\documentclass[12pt]{article}
\usepackage[textwidth=150mm, textheight=220mm]{geometry}
\usepackage{amsmath}
\usepackage{amssymb}
\usepackage{amsmath}
\usepackage{amsthm}
\usepackage{bbold}
\usepackage{xspace}
\usepackage{graphicx}
\usepackage{psfrag}
\usepackage{relsize}
\usepackage{mathrsfs}
\usepackage{hyperref}
\usepackage[bf,small]{caption2}
\setcaptionwidth{.9\textwidth}
\renewcommand{\thefootnote}{\fnsymbol{footnote}}

\numberwithin{equation}{section}

\newcommand{\be}{\begin{equation}}
\newcommand{\ee}{\end{equation}}
\newcommand{\ba}{\begin{aligned}}
\newcommand{\ea}{\end{aligned}}

\newcommand{\psu}{\mathfrak{psu}}

\def\l{{\lambda}}

\def\be{\begin{equation}}
\def\ee{\end{equation}}
\newcommand{\bea}{\begin{eqnarray}}
\newcommand{\eea}{\end{eqnarray}}
\def\D{\Delta}

\def\a {\alpha}
\def\b {\beta}
\def\s {\sigma}
\def\pa {\partial}
\def\g {\gamma}
\def\om {\omega}
\def\p{\phi}
\def\vp{\varphi}

\def\ta{\tilde{\a}}
\def\la{\label}

\def\dph{\dot{\p}}

\def\da{\dot{\alpha}}

\def\pra{\a^\prime{}}
\def\pta{\ta^\prime{}}
\def\ov{\over}

\def\E{{\rm E}}
\def\K{{\rm K}}
\def\dz{\dot{z}}
\def\H{{\cal H}}
\def\AdS{{\rm AdS}_5\times {\rm S}^5}

\def\de{\delta}
\def\eps{\epsilon}
\def\pws{p_{{\rm ws}}}
\def\j{{\cal R}}
\def\cj{{\cal J}}
\newcommand{\su}{\mathfrak{su}}

\makeatletter
\def\sla@#1#2#3#4#5{{%
  \setbox\z@\hbox{$\m@th#4#5$}%
  \setbox\tw@\hbox{$\m@th#4#1$}%
  \dimen4\wd\ifdim\wd\z@<\wd\tw@\tw@\else\z@\fi
  \dimen@\ht\tw@
  \advance\dimen@-\dp\tw@
  \advance\dimen@-\ht\z@
  \advance\dimen@\dp\z@
  \divide\dimen@\tw@
  \advance\dimen@-#3\ht\tw@
  \advance\dimen@-#3\dp\tw@
  \dimen@ii#2\wd\z@  \raise-\dimen@\hbox to\dimen4{%
    \hss\kern\dimen@ii\box\tw@\kern-\dimen@ii\hss}%
  \llap{\hbox to\dimen4{\hss\box\z@\hss}}}}
\def\slashed#1{%
  \expandafter\ifx\csname sla@\string#1\endcsname\relax
    {\mathpalette{\sla@/00}{#1}}%
  \else
    \csname sla@\string#1\endcsname
  \fi}
\makeatother

\begin{document}

%%%%%%%%%%%%%%%%%%%%%%%%%%%%%%%%%%%%%%%%%%%%%%%%%%%%%%%%%%%%%%%%%%%%%%%%%
%%%%%%%%%%%%%%%%%%%%%%%%%%%%%%%%%%%%%%%%%%%%%%%%%%%%%%%%%%%%%%%%%%%%%%%%%

\thispagestyle{empty}
\begin{flushright}\footnotesize
\texttt{hep-th/0606126}\\
\texttt{AEI-2006-049}\\
\texttt{ITP-UU-06-28}\\
\texttt{SPIN-06-24}\\
\vspace{0.8cm}
\end{flushright}

\renewcommand{\thefootnote}{\fnsymbol{footnote}}
\setcounter{footnote}{0}

\begin{center}
{\Large\textbf{\mathversion{bold}
Finite-size Effects from Giant Magnons}}\par

\vspace{1.5cm}

\textrm{Gleb Arutyunov$^{a,\dagger}$, Sergey Frolov$^{b,\dagger}$ and Marija Zamaklar$^{b}$}\vspace{8mm}

$^{a}$ {\it Institute for Theoretical Physics and Spinoza
Institute}\\
~~~~~Utrecht University,  3508 TD Utrecht, The Netherlands \\[1ex]
$^{b}$ {\it Max-Planck-Institut f\"ur
Gravitationsphysik, Albert-Einstein-Institut}\\ ~~Am M\"uhlenberg 1,
D-14476 Potsdam, Germany\\[2ex] 
{\tt \small G.Arutyunov@phys.uu.nl, frolovs@aei.mpg.de, marzam@aei.mpg.de}
\vspace{3mm}

%%%%%%%%

\par\vspace{1cm}

\textbf{Abstract}\vspace{5mm}
\end{center}

In order to analyze finite-size effects for the gauge-fixed string
sigma model on $\AdS$, we construct one-soliton solutions carrying
finite angular momentum~$J$. In the infinite~$J$ limit the solutions
reduce to the recently constructed one-magnon configuration of Hofman
and Maldacena. The solutions do not satisfy the level-matching
condition and hence exhibit a dependence on the gauge choice, which
however disappears as the size~$J$ is taken to
infinity. Interestingly, the solutions do not conserve all the global
charges of the $\psu(2,2|4)$ algebra of the sigma model, implying that
the symmetry algebra of the gauge-fixed string sigma model is
different from $\psu(2,2|4)$ for finite~$J$, once one gives up the
level-matching condition. The magnon dispersion relation exhibits
exponential corrections with respect to the infinite~$J$ solution.  We
also find a generalisation of our one-magnon configuration to a
solution carrying two charges on the sphere.  We comment on the
possible implications of our findings for the existence of the Bethe
ansatz describing the spectrum of strings carrying finite charges.

\vspace*{\fill}

$^{\dagger}$ {\small Correspondent fellow at Steklov Mathematical Institute,
Moscow }

\newpage
\setcounter{page}{1}
\renewcommand{\thefootnote}{\arabic{footnote}}
\setcounter{footnote}{0}

\tableofcontents
 \newpage
%%%%%%%%%%%%%%%%%%%%%%%%%%%%%%%%%%%%%%%%%%%%%%%%%%%%%%%%%%%%%
\section{Introduction and summary}

Recent studies of string theory in $\AdS$ and the dual ${\cal N}=4$
super Yang-Mills theory, motivated by the AdS/CFT duality conjecture
\cite{M}, have led to new interesting insights into the problem of
finding the spectrum of quantum strings in the $\AdS$ geometry. It
seems that this complicated problem can be addressed in two stages.
String states can be naturally characterized by the charges they carry
under the global symmetry algebra of the $\AdS$ space-time.  In the
first stage one considers states for which one of the angular momenta
on the five-sphere is infinite. In this case the problem of finding
and classifying the corresponding string states simplifies
considerably.  In the second stage, it may then be possible to
bootstrap this analysis to string states with finite charges. 

Perhaps the easiest way to appreciate the simplifying features of the
infinite-charge limit is to consider the light-cone gauge-fixed string
theory. In the light-cone gauge (for a precise definition see
section~2) the gauge fixed world-sheet action depends explicitly on
the light-cone momentum, which can be thought of as one of the global
symmetry charges. By appropriately rescaling a world sheet-coordinate,
the theory becomes defined on a cylinder of circumference proportional
to the value of the light-cone momentum.  At this stage, one can
consider the decompactifying limit, i.e.~the limit in which the radius
of the cylinder goes to infinity while keeping the string tension
fixed \cite{MP}-\cite{HM}.  In this limit one is left with the theory
on a plane which leads to significant simplifications. In particular,
the notion of asymptotic states is well defined.  Furthermore, since
the light-cone gauge fixing manifestly breaks conformal invariance,
the world-sheet theory has a massive spectrum. This theory is
(believed to be) integrable at the quantum level, and hence a
multi-body interaction factorises into a sequence of two-body
interactions.\footnote{While integrability is known to be broken
beyond the planar level, it seems to be preserved if one focuses on
the specific set of most probable string splitting channels
\cite{split}.}  Thus the problem of solving the theory basically
reduces to the problem of finding the dispersion relation for
elementary excitations and the two-body
S-matrix. These two quantities have not as yet been determined from
the first principles of field theory. However, the insights coming
from gauge theory \cite{BMN}-\cite{Beisert:2004ry} from semi-classical
string quantisation \cite{BMN, GKP}-\cite{FPZ} as well as from the
analysis of classical strings \cite{Bena:2003wd}-\cite{Das:2004hy}
lead to a conjecture for the form of the dispersion relation and the
corresponding S-matrix \cite{BDS, AFS}.  From the perspective of
relativistic field theory, both the dispersion relation and the
S-matrix have an unusual form. The dispersion relation has been
conjectured to be
\begin{equation}
\label{disp}
\epsilon(p) = \sqrt{1 + \frac{\lambda}{\pi^2} \sin^2 \frac{p}{2}} \, .
\end{equation}
The appearance of the $\sin p/2$ in the dispersion relation is a
common feature of theories on a lattice, but its origin from the
world-sheet perspective remains obscure, given that the string
world-sheet is continuous. Secondly, the dispersion relation is
not Lorentz invariant.
This is basically a consequence of the gauge fixing which manifestly
breaks Lorentz invariance.  Yet, the dispersion relation is of
relativistic form (it has a square root) signaling the possibility of
having ``anti-particles'' in the theory, corresponding to a different
choice of the sign in front of the square root.

The structure of the S-matrix was initially proposed
in~\cite{AFS,S,BS,FPZ} based on an ``empirical'' analysis of
semiclassical string spectra. It turns out however~\cite{Dynam},
that the structure of the S-matrix is uniquely fixed by the global
$\su(2|2) \times \su(2|2) \subset \psu(2,2|4)$ symmetry, up to an
unknown scalar function $\sigma(p_1,p_2)$, the so-called dressing
factor. Ideally, one would hope that further physical
requirements, such as unitarity, factorization and additional
symmetries of the theory would uniquely fix this factor. In
particular, in \emph{relativistic} integrable quantum field
theories, implementation of Lorentz invariance is particularly
constraining. It introduces an extra equation, the crossing
relation, that the S-matrix has to satisfy \cite{ZZ}. This
equation relates the S-matrix that scatters particles with the
S-matrix that scatters particles with antiparticles, and basically
has a unique solution (with the minimal number of poles/zeros in
the physical region).

Unfortunately, the light-cone gauge-fixed sigma model is not
Lorentz invariant, and this is explicitly reflected in the Lorentz
non-invariant form of the S-matrix: it depends separately on the
magnon rapidities, rather than on their difference. However, it
was argued in~\cite{Janik} that ``traces'' of Lorentz invariance
should be present in this model and that some version of the
relativistic crossing relation should hold for the S-matrix in
this model. Using the Hopf-algebraic formulation of crossing in
terms of an antipode, a functional equation for the dressing
factor was derived in~\cite{Janik}.

The dressing factor $\s$ explicitly depends on the coupling
$\sqrt{\lambda}$, and it admits a ``strong coupling'',
$1/\sqrt{\lambda}$ expansion. Currently, the first two orders in the
expansion have been computed in \cite{AFS,HL,Frey}, building upon
observations of \cite{SZZ,BT,SZ}. It was demonstrated in
\cite{AF06} that, up to this order, the dressing factor indeed satisfies
the functional equation of \cite{Janik}. It remains an important open
problem to find the solution to this equation. It appears however,
that the solution is not unique, and that additional physical
constraints need to be imposed~\cite{SJ}.

At large $\l$ the problem of deriving the dispersion
relation~(\ref{disp}) and the string S-matrix can be addressed in
the classical string sigma model, as was recently pointed out in
\cite{HM}. It was shown there that in \emph{the decompactifying
limit} a one-magnon excitation with finite world-sheet momentum
$p$ can be identified with a one-soliton solution of the classical
string sigma model. The corresponding string configuration carries
infinite energy and infinite angular momentum $J$, since it
describes the theory on a plane. The difference of the two is,
however, finite and equal to the energy of the world-sheet
soliton; it is $\sqrt{\lambda}/\pi |\sin(p/2)|$ which is precisely
the large $\lambda$ limit of the dispersion relation~(\ref{disp}).
A single-magnon excitation obviously does not correspond to a
physical configuration of the \emph{closed string} since it
carries a non-vanishing world-sheet momentum.  The subtle point in
the consideration of \cite{HM} is that to describe these magnons
in the light-cone gauge-fixed sigma model, one has to give up the
level-matching condition.  This implies that the corresponding
target space string configuration, is an \emph{open, rigidly
moving} string, such that the distance between the string
endpoints is constant in time and is proportional to the
world-sheet momentum of the magnon. In the \emph{conformal gauge}
supplemented by the condition $t=\tau$ this translates into
nontrivial boundary conditions on the space coordinate appearing
in the light-cone coordinates. A configuration with these
characteristics was then constructed as a sigma model solution in
the conformal gauge, and named the giant magnon \cite{HM}.

\medskip
\medskip

In whatever way one solves the theory on the plane, an important
problem one has to face afterwards is how to ``upgrade'' the
findings from a plane to a cylinder. All physical string
configurations are characterised by a \emph{finite} value of the
light-cone momentum, and as such they are excitations of a theory
on a cylinder rather then on a plane.  In this paper we try to
systematically address the question of what kind of modification
finite size effects can introduce.

In general going from a theory on a plane to a theory on a
cylinder may modify the theory significantly. While on a plane it
is always possible to construct a multi-particle state as a
superposition of well-separated single-particle excitations, this
is no longer the case once we are on a cylinder. However, if the
size of the cylinder~$L$ is very large, much larger than the size
of the excitation and much larger than the range of the
interactions, then the leading finite-size effects could be
incorporated through the following asymptotic construction.  The
dispersion relation for a single excitation is taken to be the
same as in the infinite volume system, the energy of a
multi-particle system is taken to be additive, and the structure
of the wave function is unmodified. The only way in which
finite-size effects modify the consideration from the plane, is
via periodic boundary conditions which eigenstate wave functions
have to satisfy. In the case of a spin chain, the boundary
conditions on the wave function basically lead to Bethe equations.
In some cases, like for example for the XXX spin chain, this
asymptotic construction remains exact for any size of the
finite-size system. However, for spin chains with long-range
interactions, such are those which arise in higher-order
perturbation gauge theory, the asymptotic construction is valid
only for long spin chains.  Once the range of interactions between
magnons becomes of the size of the system, the asymptotic
construction has to be modified, and finite-size effects (the
wrapping interactions in the gauge theory language) have to be
taken into account.

\medskip
\medskip

In this paper, we address finite size effects at large $\l$, that is
in the classical string sigma model.  An analysis of these effects by
doing a semi-classical string computation is generically very
involved, since a typical closed string state is a complicated
superposition of a large number of elementary, magnon
excitations. However, the consideration simplifies drastically if,
following \cite{HM}, one gives up the level-matching condition and
considers a single magnon. Giving up the level-matching condition at
finite values of the light-cone momentum is actually what one always
does in the process of quantising light-cone gauge-fixed string theory
in flat space. The level matching is imposed only at the very last
stage of quantisation.  In this paper, we do it in the classical
theory, and construct solutions of the sigma model corresponding to a
single magnon excitation for the theory on a cylinder. We find magnon
solutions of the string sigma model in the conformal gauge and in a
one-parameter family of light-cone gauges, labeled by a parameter~$a$,
\begin{equation}
\label{lca}
x_+ = (1-a) t + a \phi = \tau \, ,  \quad x_-  = \phi - t \, , \quad
p_+ = (1-a)\,  p_\phi - a\, p_t = \text{const} \, .
\end{equation}
Many new features appear with respect to the case of infinite volume.
The first and probably the most striking result at first glance, is
that a magnon in the finite size system is a \emph{gauge dependent
object}: its target space picture and the dispersion relation
explicitly depend on the parameter~$a$. Furthermore, all of these
various magnon configurations reduce to \emph{the same} configuration
in the limit of infinite light-cone momentum, i.e~in the limit where
the size of the system is taken to infinity.  Both of these results
however should not come as a surprise. Namely, one way of
``constructing'' a single magnon configuration is to start with the
physical, closed string state which describes the system of two
magnons (with vanishing total worldsheet momentum). To isolate a
one-magnon state, we need to cut this closed string, and separate the
magnons from each other.  In principle, cutting of the string is an
unphysical process, since its obviously breaks reparametrisation
invariance (it declares that different parts of the string are
physically different). Hence cutting of the string may introduce gauge
dependence, depending on how we decide to open the string. A natural
way of opening the string is dictated by dropping the level matching
condition. In the light-cone gauges~(\ref{lca}) it implies that
\begin{equation}
\Delta x_- = - \int_{-\frac{\pi}{\sqrt{\lambda}} P_+ }^{\frac{\pi}{\sqrt{\lambda}} P_+}
d \sigma p_i x^i{}' \, \neq 0 ,
\end{equation}
where $P_+$ is the total light-cone momentum and $x^i$ and $p_i$ are
transverse coordinates and momenta. In other words, if level matching
is not satisfied in the gauge labeled by $a$, the string opens in the
$x_-$ direction, so that the separation of its endpoints in this
direction is constant with respect to the time $x^+$. Note however,
that the derivatives of the transverse fields $x^i$ \emph{do not
vanish} at the string endpoints, and hence correspondingly the
world-sheet momentum does not vanish there.  The world-sheet momentum
is however conserved, as a consequence of the periodic boundary
conditions which $p_i, x^i$ satisfy. In other words, although the
world-sheet momentum ``flows out'' of the string on one side, it
``flows in'' from the other side, due to the periodic boundary
conditions.

\begin{figure}[t]
\begin{center}
\includegraphics*[width=.7\textwidth]{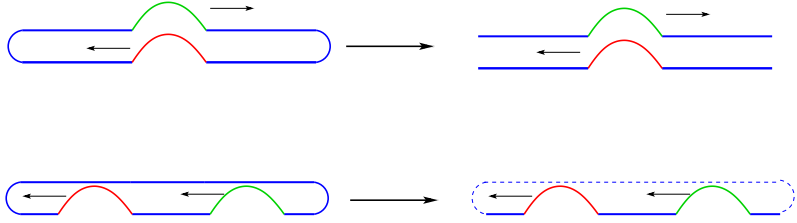}
\end{center}
\caption{\label{limi}There are potentally two ways to take the limit from a finite
 $J$, two soliton configuration.
One way is to have the solitons on
 ``different sides'' of the string: this leads to two
 one-soliton configurations, living on different lines. Another way is
 to have the solitons on the same ``side'' of the string: this leads
 to a nontrivial two-soliton configuration on the line.  In the target
 space, the former configuration corresponds to a folded string with
 the shape of a giant magnon, which is a legitimate closed string
 state. In the latter case, sending $J$ to infinity, does not
 naturally opens up the string, since solitons remain unseparated
 in the limit. Only if the total worldsheet momentum is nonzero, the
 latter  becomes a complicated open string state, which is such that when
 the total worldsheet momenta of solitons becomes zero, one is back to
 the closed string.  }
\end{figure}

In the infinite volume case the situation is physically different,
since the transverse fields satisfy both periodic and \emph{Neumann boundary
conditions}, making thus the one-magnon case closer to a conventional open
string state . This is also manifested in the fact that a proper closed
string state can be build trivially out of two infinite-$J$ magnons,
by putting them on top of each other (one ends up with a folded closed
string which is made out of two copies of the giant magnons
of~\cite{HM}), see figure~\ref{limi}. This should be contrasted with
the finite-$J$ situation where one
cannot build a closed string state by trivially putting two
one-magnons on top of each other.

Although our one-magnon configurations are gauge dependent, the
requirement that the spectrum of physical excitations is gauge
independent imposes severe constraints on the structure of the
theory. It is plausible that for finite $J$, there is a preferred
choice of the parameter~$a$ simplifying the exact quantisation of the
model.
Our analysis indicates that it would be the temporal gauge,
$t=\tau\,,\ p_\p=J$, corresponding to $a=0$.
The suggestive reason for this is that, as we show in this paper,
only for the $a=0$ gauge
one can identify the world-sheet momentum~(\ref{pws}) with the
spin-chain magnon momentum.

\begin{figure}[t]
\begin{center}
\includegraphics*[width=.7\textwidth]{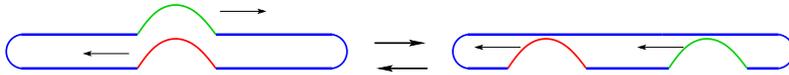}
\end{center}
\caption{\label{First-sol}At finite $J$ the two-soliton configuration is complicated
and never a trivial superposition of two one-magnon
solutions. This is the reason why we cannot \emph{trivially} build a closed
string state only from two magnons. At infinite $J$ the situation is
different, and there is a trivial configuration of two magnons (see
the upper right-hand side picture of figure~\ref{limi}).
}
\end{figure}

The second result of our analysis is that the dispersion relation for
the one-magnon case receives \emph{exponential corrections} with
respect to the infinite~$J$ case.
\bea\nonumber
E-J=\frac{\sqrt{\lambda}}{\pi}\sin\frac{\pws}{2}
\Big(1-\frac{4}{e^2}\sin^2\frac{\pws}{2}~ e^{-\j} +\cdots\Big)\,,
\eea
where $\j$ is the effective length felt by the magnon with momentum $\pws$
\bea\nonumber
\j =\frac{2\pi J}{\sqrt\l\sin\frac{\pws}{2}} +a\pws\cot\frac{\pws}{2}\,.
\eea
This formula  shows explicitly a nontrivial dependence on the parameter $a$. Moreover,
the dispersion relation is  periodic in $\pws$ only
for $a=0$. This is the reason why the $a=0$ gauge seems to be preferred from a gauge theory
perspective.

It is known that the one-magnon configuration is
half-supersymmetric, i.e. the energy of the magnon~(\ref{disp})
is determined by the BPS relation~(\ref{disp}) which follows from
the centrally extended $\su(2|2)\times \su(2|2)$ algebra
\cite{Dynam, HM}. Still, the magnon energy receives finite-size corrections, and this
implies that the central charge in the algebra should also receive
finite-size corrections.

This brings us to the third result of our analysis. Namely, by
explicitly evaluating the charges of the ${\rm SO}(3)$ algebra on
our one-magnon configurations, one can check that the off-diagonal
charges \emph{are not} preserved in time.\footnote{ 
Rephrasing F.~Dostoevsky, we could summarise our findings in one sentence:~``If
there is no 

~~~~God, everything is broken''.} As we explicitly show
(see section~\ref{ch}) this is a simple consequence of the fact
that one dismisses the level matching condition, and of the fact
that the transverse fields do not satisfy Neumann boundary
conditions. If $J$ is infinite, all charges \emph{are conserved}
since the open string satisfies standard boundary conditions.

The breaking of the algebra may sound worrisome. A similar
phenomenon has however already appeared in the case of the
asymptotic all-loop Bethe ansatz in~\cite{BS}, where only after
imposing the momentum conservation one recovered the full
$\psu(2,2|4)$ algebra.\footnote{We thank Matthias Staudacher for
this comment.}  Also, the algebraic construction of the S-matrix
in~\cite{Dynam} involved only the $\su(2|2) \times \su(2|2)$
subalgebra, rather then the full $\psu(2,2|4)$ algebra. It would
be very important to understand the structure of the finite $J$,
off-shell algebra more explicitly.

In the last section we generalise our finite $J$ magnon configuration
to the case of two spins. In the infinite $J$ limit this configuration
reduces to the two-spin giant magnon solution of~\cite{Dorey}. Our
method to obtain this solution is however different from the one used
in~\cite{Dorey} and may be more applicable for the construction of the
three-spin configuration.

\begin{figure}[t]
\includegraphics[width=.245\textwidth]{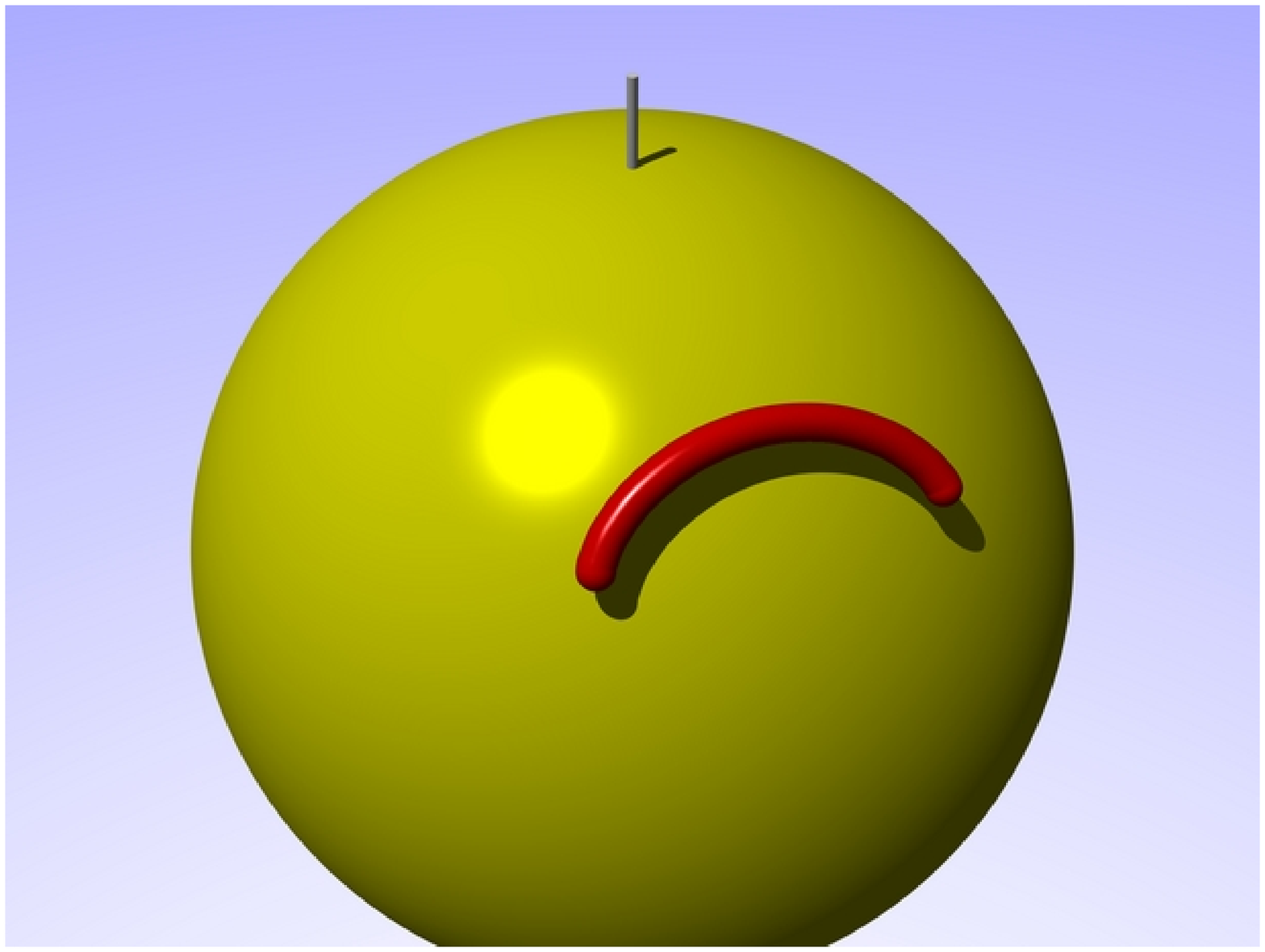}
\includegraphics[width=.245\textwidth]{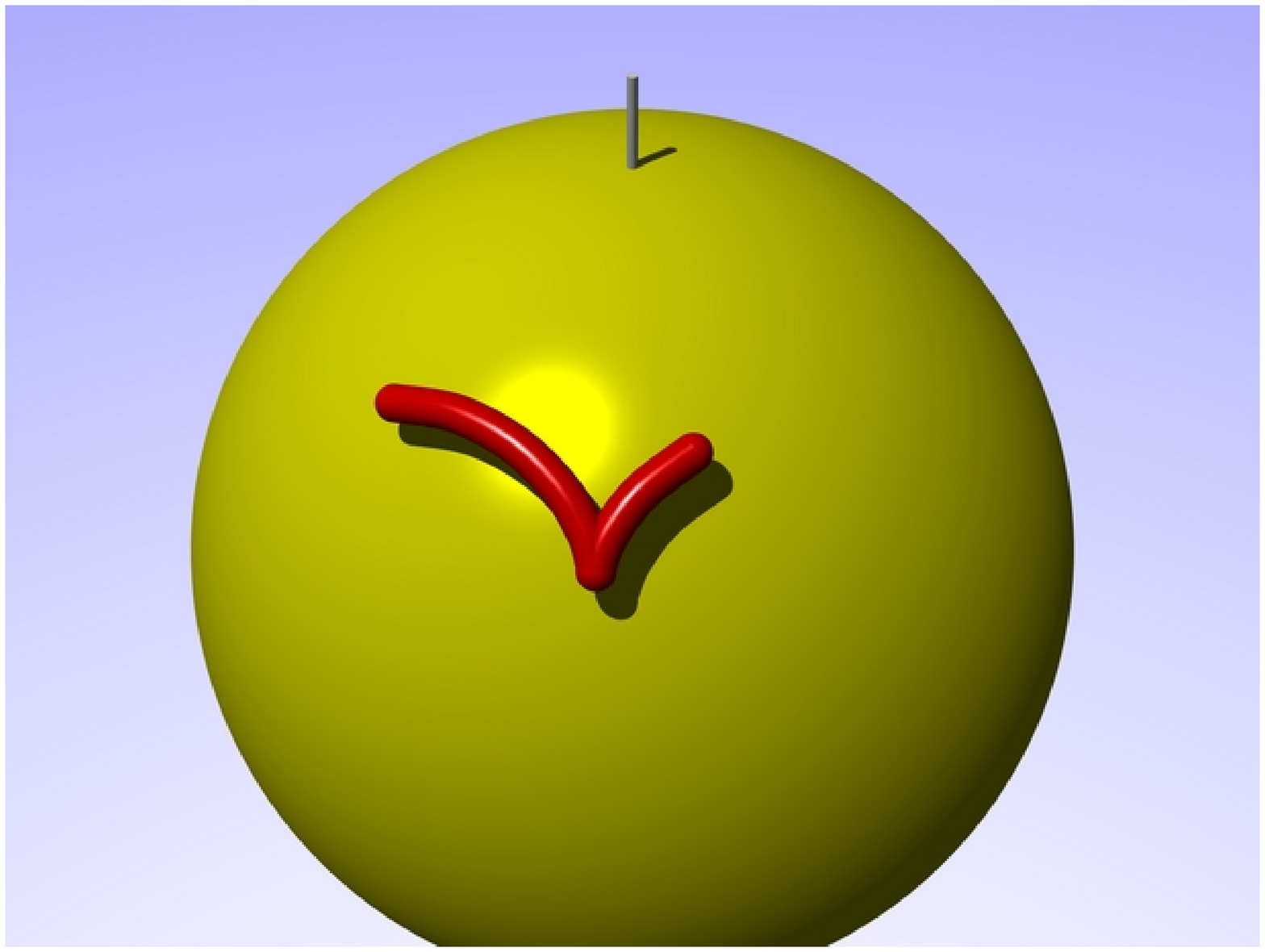}
\includegraphics[width=.245\textwidth]{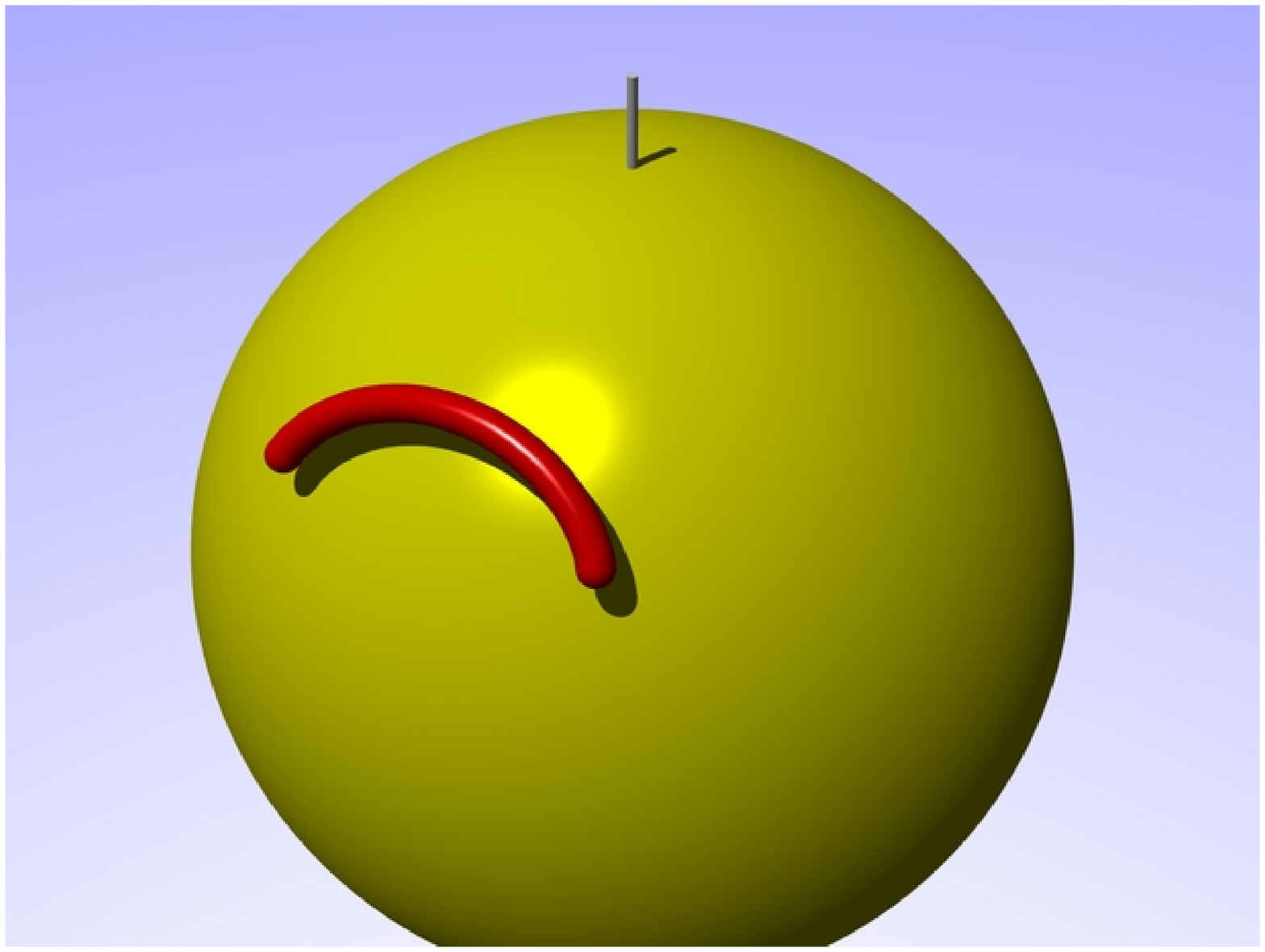}
\includegraphics[width=.245\textwidth]{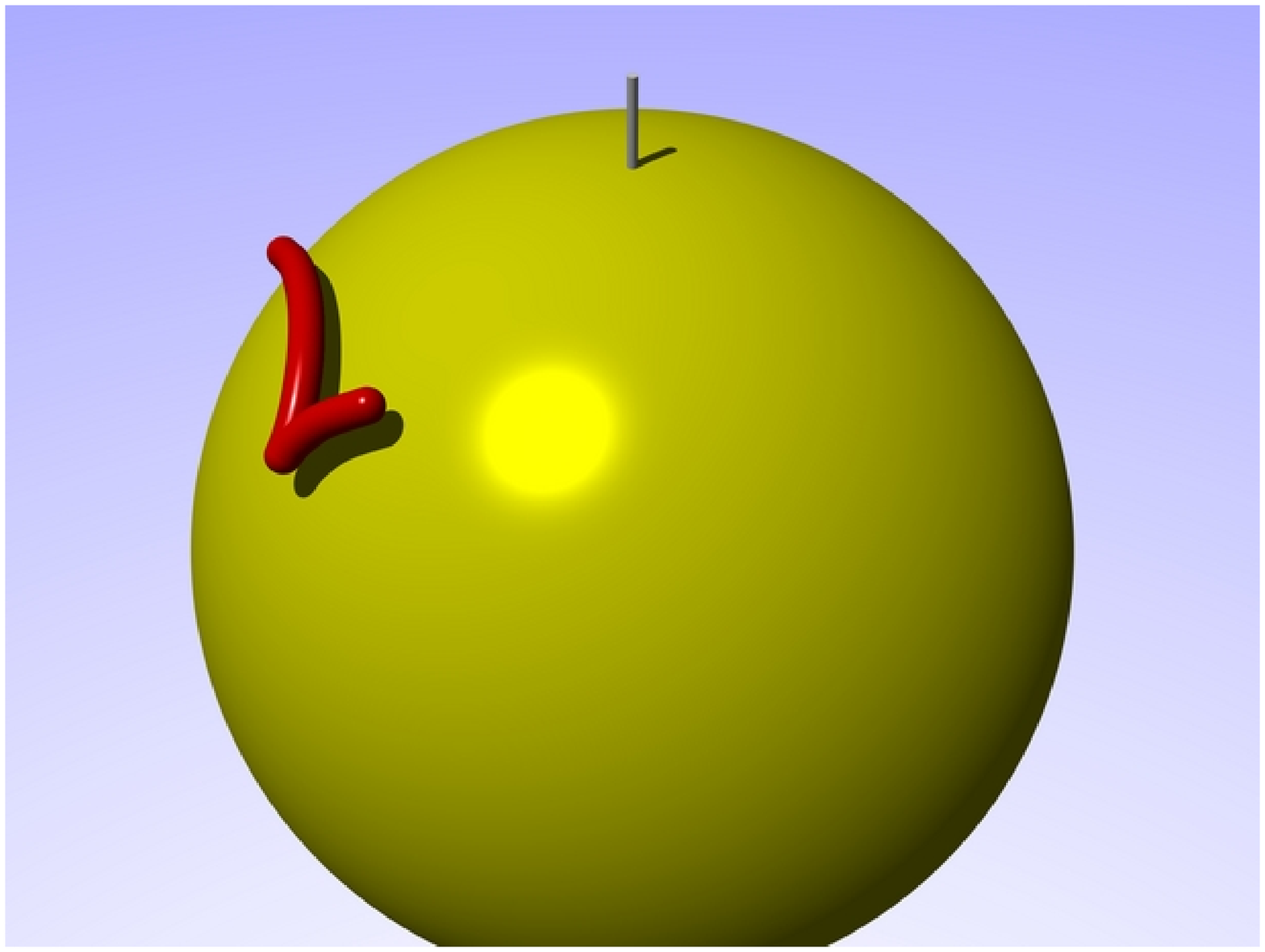}
\caption{\label{snapshots}Snapshots of the time evolution of the solution in conformal gauge.}
\end{figure}
Finally, let us comment on the target space picture of the finite-size
giant magnon. A simulation of the time evolution of the solution in
the conformal and $a=0$ gauges can be seen
at~\url{http://www.aei.mpg.de/~peekas/magnons/}. Several snapshots
from this movie are shown in figure~\ref{snapshots}. We see that,
unlike the infinite $J$ magnon, the string configuration is
\emph{nonrigid}: propagation of the soliton from one end of the string
to the other happens in a finite target space time, and leads to
``wiggly'' behavior of the string. The boundaries of the ``stripe'' in
which the string wiggles depend on $J$ and the world-sheet momentum
$p$.  There are three limits which one can take, and which link our
solution to the known string configurations. These thus serve as a
crosscheck of the solution.  First, as $J \rightarrow \infty$ the
solution reduces to the solution of~\cite{HM}.  The string endpoints
touch the equator and the period with which the string wiggles goes to
infinity. Secondly, if we keep $J$ finite, and send the world-sheet
momentum to its maximal value $p=\pi$, the solution reduces to half of
the rigid folded string of~\cite{GKP}. Finally, as the world-sheet
momentum is sent to zero, the giant magnon
``shrinks'' to zero size, and reduces to a massless point particle
moving on the equator with the angular momentum equal to $J$.

We end the introduction with a summary of the potential implications
of our findings to the Bethe ansatz approach to the quantisation of
strings. The crucial feature necessary for the formulation of the
Bethe ansatz, is additivity of the energy for the multi-magnon
excitations. This feature seems to be lost in the case of the finite
$J$ configurations.  Namely, giant magnons correspond to world-sheet
solitons. Typically multi-soliton configurations are not simple
superpositions of one-soliton configurations, neither for finite nor
for infinite spaces.  In addition, in finite volumes the very
definition of soliton number becomes obscure.  Yet, in our case it
seems that there exists a (at least heuristic) way of counting the
number of solitons present in the finite $J$ string. These should
correspond to the number of (target space) spikes characterising the
string configuration. However, as explained in section \ref{ch} the
energy of such a multi-magnon configuration would not generically be
realised as the sum of the energies of one-magnon configurations,
carrying the appropriate fraction of the total charge.  This seems to
imply that at finite $J$ the string spectrum would not be described by
a simple Bethe ansatz of the form~\cite{AFS}. If the Bethe ansatz
description of the string spectrum at finite $J$ is at all possible,
then it is plausible that this would require the introduction of
auxiliary excitations similar to the constructions of \cite{MP,RSS,
GKSV,Gromov:2006cq, Mann}.

%%%%%%%%%%%%%%%%%%%%%%%%%%%%%%%%%%%%%%%%%%%%%%%%%%%%%%%%%%%%%
%%%%%%%%%%%%%%%%%%%%%%%%%%%%%%%%%%%%%%%%%%%%%%%%%%%%%%%%%%%%%
\section{String theory in a uniform gauge }
%%%%%%%%%%%%%%%%%%%%%%%%%%%%%%%%%%%%%%%%%%%%%%%%%%%%%%%%%%%%%
In this section we review a class of uniform gauges for
strings propagating on a target manifold. These gauges generalize
the standard phase-space light-cone gauge of \cite{Goddard:1973qh}
to a curved background, and have been used to study the dynamics of strings in
$\AdS$ \cite{Metsaev:2000yu,KT,AF,AAT,AAF,AFlc,FPZ}. In our discussion we follow closely
\cite{AF,AFlc}.
\medskip

We consider strings propagating on a target manifold possessing (at least)
two abelian isometries realized by shifts of
the time coordinate of the manifold denoted by $t$, and a space coordinate
denoted by $\phi$. If the variable $\phi$ is an angle then
the range of $\phi$ is from $0$ to $2\pi$.

\medskip

To impose a uniform gauge
we also assume that the string sigma-model action is invariant
under shifts of  $t$ and
$\phi$, with all the other bosonic and fermionic fields being
invariant under the shifts. This means that the string action does
not have an explicit dependence on $t$ and $\p$ and depends only
on the derivatives of the fields. An example of such a string
action is provided by the Green-Schwarz superstring in $\AdS$
where the metric can be written in the form
\bea \nonumber ds^2 =
G_{tt}\, dt^2\, +\, G_{\p\p}\, d\p^2\, +\, G_{ij}\,dx^idx^j \, .
\eea
Here $t$ is the global time
coordinate of $\rm{AdS}_5$, $\p$ is an angle of ${\rm S}^5$, and
$x^i$ are the remaining 8 coordinates of  $\AdS$.
Strictly speaking, the original
Green-Schwarz action presented in \cite{MT} contains fermions
which are charged under the U(1) transformations generated by the
shifts of $t$ and $\p$. However, it is possible to redefine the
fermions and make them neutral, see
\cite{AAF, AAFg} for details.

\medskip

To simplify the notations we consider explicitly only the bosonic
part of a string sigma model action, and assume that the B-field
vanishes. A most general fermionic Green-Schwarz action can be
analyzed in the same fashion, and leads to  the same conclusions.
The corresponding part of the string action  can be written in the
following form \bea \la{S1} S = -{\sqrt{\lambda}\over
4\pi}\int_{-r}^{ r}\, {\rm d}\s{\rm d}\tau\, \g^{\a\b}\partial_\a
X^M\partial_\b X^N\, G_{MN}\,. \eea 
Here ${\sqrt{\lambda}\over
2\pi}$ is the effective string tension, which for strings in $\AdS$ 
is related to the radius of ${\rm S}^5$ as $\sqrt{\lambda} =
R^2/\a'$. Coordinates $\s$ and $\tau$ parametrize the string
world-sheet. For later convenience we assume the range of $\s$
to be $- r\le \s\le  r$, where $r$ is an arbitrary constant. The
standard choice for a closed string is $r=\pi$. Next,
$\g^{\a\b}\equiv \sqrt{-h}\, h^{\a\b}$ is a Weyl-invariant
combination of the world-sheet metric $h^{\a\b}$ which  in the conformal
gauge is equal to $\g^{\a\b} = {\mbox{diag}}(-1,1)$. Finally, $X^M =
\{t,\p,x^i\}$ are string coordinates and $G_{MN}$ is the
target-space metric which is independent of $t$ and $\p$.

\medskip

The simplest way to impose a uniform gauge is to introduce momenta
canonically-conjugate to the coordinates $X^M$ \bea p_M = {2\pi\ov
\sqrt\l}{\de S\ov \de \dot{X}^M} = -\g^{0\b}\partial_\b X^N\,
G_{MN}\,,\quad \dot{X}^M\equiv \pa_0 X^M \,, \eea and rewrite the
string action (\ref{S1}) in the first-order form \bea \la{S2} S
={\sqrt{\lambda}\over 2\pi} \int_{- r}^{ r}\, {\rm d}\s{\rm d}\tau\,
\left( p_M \dot{X}^M + {\g^{01}\ov\g^{00}} C_1+ {1\ov 2
\g^{00}}C_2\right)\,.  \eea 
The reparametrisation invariance of the
string action leads to the two Virasoro constraints 
\bea \nonumber
C_1=p_MX'^M\,,\quad C_2=G^{MN} p_M p_N + X'^M X'^N G_{MN}\,,\qquad
X'^M\equiv \pa_1 X^M\,, \eea 
which are to be solved after imposing a gauge condition.

The invariance of the string action under the shifts leads to the
existence of two conserved charges
\bea\la{charges}
E = - {\sqrt{\lambda}\over 2\pi}
\int_{- r}^{ r}\, {\rm d}\s\, p_t\ \, , \qquad J={\sqrt{\lambda}\over 2\pi}
\int_{- r}^{ r}\, {\rm d}\s\, p_\p\ .
\eea
It is clear that the charge $E$ is the target space-time energy
and $J$ is the total U(1) charge of the string.

%\medskip
To impose a uniform gauge we introduce the ``light-cone'' coordinates
and momenta: \bea \la{lcc} &&x_- =\p \,-\,t\ , \,\, x_+ =(1-a)\,t\,
+\,a\,\p\ ,\,\, p_- = p_\p\,+\,p_t\ ,\,\, p_+ = (1-a) p_\p
\,-\,a\,p_t\ ,\nonumber \\ &&t = x_+ \,-\,a\,x_-\ ,\, \, \p = x_+
\,+\, (1-a) x_-\ , \,\, p_t = (1-a)\,p_- - p_+ \ ,\,\, p_\p = p_+
\,+\, a\,p_- \ . \nonumber \eea Here, $a$ is an arbitrary number which
parametrizes the most general uniform gauge up to some trivial
rescaling of the light-cone coordinates such that the light-cone
momentum $p_-$ is equal to $p_\p\,+\,p_t$. This choice of gauge is
natural in the AdS/CFT context because, as we will see in a moment, in
a uniform gauge the world-sheet Hamiltonian is equal to $E-J$.  Taking
into account (\ref{charges}), we get the following expressions for the
light-cone charges \bea\la{charges2}\nonumber P_-
\,=\,{\sqrt{\lambda}\over 2\pi} \int_{- r}^{ r}\, {\rm d}\s\, p_-\,=
\, J\,-\,E\ \, , \qquad P_+ \,=\,{\sqrt{\lambda}\over 2\pi} \int_{-
r}^{ r}\, {\rm d}\s\, p_+\,=\, (1-a)\, J \,+\, a\,E\, . \eea In terms
of the light-cone coordinates the action (\ref{S2}) takes the form
\bea \la{S3} S ={\sqrt{\lambda}\over 2\pi} \int_{- r}^{ r}\, {\rm
d}\s{\rm d}\tau\, \left( p_-\dot{x}_++ p_+ \dot{x}_- + p_i \dot{x}^i +
{\g^{01}\ov\g^{00}} C_1+ {1\ov 2 \g^{00}}C_2\right)\, , \eea where
\bea \la{C1} C_1\,=\,p_+x_-' \,+\, p_-x_+' \,+\, p_ix'^i\,, \eea and
the second Virasoro constraint is a quadratic polynomial in $p_-$.

We then fix the uniform light-cone gauge by imposing the
conditions \bea \la{ulc} x_+ \,=\, \tau \,+\, a m\s\ ,\quad\quad
p_+ \,=\, 1\ . \eea The integer number $m$ is the winding number
which appears because the coordinate $\p$ is an angle variable
with the range $0\le\p\le 2\pi$. The consistency of this gauge
choice forces us to choose the constant $r$ to be \bea r =
{\pi\ov\sqrt{\lambda}}P_+\,. \eea To find the gauge-fixed action,
we first solve the Virasoro constraint $C_1$ for $x_-'$ \bea
C_1\,=\,x_-' \,+\,  a m p_- \,+\, p_ix'^i\,=\,0\, \quad
\Longrightarrow \quad x_-'\,=\, - a m p_- \,-\, p_ix'^i\,, \eea
substitute the solution to $C_2$ and solve the resulting
quadratic equation for $p_-$. Substituting all these solutions
into the string action (\ref{S3}), we end up with the gauge-fixed
action \bea \la{S4} S ={\sqrt{\lambda}\over 2\pi} \int_{- r}^{
r}\, {\rm d}\s{\rm d}\tau\, \left( p_i \dot{x}^i \,-\, \H
\right)\, , \eea where \bea \la{denH} \H \,=\, -p_-(x^i,x'^i) \eea is the
density of the world-sheet Hamiltonian which depends only on the physical
(transverse) fields $x^i$. It is worth noting that $\H$ has no
dependence on $\l$, and the dependence of the gauge-fixed action
on $P_+$ comes only through the integration limits $\pm r$.

\medskip

The world-sheet Hamiltonian in this gauge is related to the target space-time energy $E$ and the U(1) charge $J$ as follows
\bea
\la{H}
H \,=\, {\sqrt{\lambda}\over 2\pi}
\int_{- r}^{ r}\, {\rm d}\s\, \H = E-J\,.
\eea
In the
AdS/CFT correspondence the space-time energy $E$ of a string state
is identified with the conformal dimension $\D$ of the dual CFT
operator: $E\equiv \D$. Since the Hamiltonian $H$ is a function of
$P_+=(1-a)J + aE$, for generic values of $a$ the relation (\ref{H}) gives us a nontrivial equation
on the energy $E$. Computing the spectrum of $H$ and solving the
equation (\ref{H}) would allow us to find conformal dimensions
of dual CFT operators.

\medskip

There are three natural choices of the parameter $a$. If $a=0$ we get
the temporal gauge $t = \tau\,,\ p_+ = J$. For strings moving in the
${\mathbb R}\times {\rm S}^5$ subspace of $\AdS$ this gauge choice is
related to the conformal gauge supplemented by the condition $t=\tau$
we use in section \ref{cg} to find the finite $E$ one-magnon
configuration. It was shown in \cite{AF} that this gauge was
(implicitly) used in \cite{Callan} to compute $1/J$ corrections in the
near BMN limit \cite{BMN}, and that the gauge-fixed Hamiltonian
describes an integrable model. It is clear that the consideration of
\cite{AF} can be straightforwardly generalized to any $a$ and
therefore, for fixed $\l\,, \ P_+\,,\ m$, the gauge-fixed Hamiltonians
define a one-parameter family of integrable models.  If $a={1\ov 2}$,
we obtain the uniform light-cone gauge $x_+ ={1\ov 2}(t +\p) =\tau\,,\
P_+ = {1\ov 2}(E+J)$ introduced and used in \cite{AFlc} to analyze the
$\su(1|1)$ subsector (see also \cite{Met}). The light-cone gauge
appears to simplify drastically computations of $1/J$ corrections in
the near BMN limit as was demonstrated in \cite{AFlc,FPZ}. It also
allows to reformulate the quantum string Bethe ansatz \cite{AFS} in a
simpler form \cite{FPZ}.  Finally, one can also set $a=1$. In this
case, the uniform gauge reduces to $x_+=\p =\tau\,,\ P_+=E$, where the
angle variable $\p$ identified with the world-sheet time $\tau$, and
the energy $E$ distributed uniformly along the string. String theory
in $\AdS$ has not been analyzed in this gauge yet.

\medskip

Since we consider closed strings, the transverse fields $x^i$ are
periodic: $x^i(r) = x^i(-r)$. Therefore, the gauge-fixed action
defines a two-dimensional model on a cylinder of circumference $2 r
={2\pi\ov\sqrt{\lambda}}P_+$.  In addition, the physical states should
also satisfy the level-matching condition \bea\la{LM} \D x_-= \int_{-
r}^{ r}\, {\rm d}\s x_-' ={2\pi\ov \sqrt\l} am H - \int_{- r}^{ r}\,
{\rm d}\s p_ix'^i = 2\pi m\,.  \eea The gauge-fixed action is
obviously invariant under the shifts of the world-sheet coordinate
$\s$.  This leads to the existence of the conserved charge
\bea\la{pws} p_{{\rm ws}} = -\int_{- r}^{ r}\, {\rm d}\s p_ix'^i\,,
\eea which is just the total world-sheet momentum of the string. In
what follows we will be interested in the zero-winding number case,
$m=0$.  Then the level-matching condition just says that the total
world-sheet momentum vanishes for physical configurations \bea\la{LM2}
\D x_-= p_{{\rm ws}} =0\,,\quad m=0\,.  \eea

\medskip

The gauge-fixed action can be used to analyze string theory in various
limits. One well-known limit is the BMN limit \cite{BMN} in which one
takes the $\l\to\infty$ and $P_+ \rightarrow \infty$ while keeping
$\tilde{\l}=\l/P_+^2$ fixed. In this case it is useful to rescale $\s$
so that the range of $\s$ would be from $-\pi$ to $\pi$. The
gauge-fixed action admits a well-defined expansion in powers of
$1/P_+$, with the leading part being just a quadratic action for 8
massive bosons (and 8 fermions). The action can be easily quantized
perturbatively, and used to find $1/P_+$ corrections
\cite{Callan,FPZ}.

\medskip

Another interesting limit is the decompactifying limit where
$P_+\to\infty$ with $\l$ kept fixed.  In this limit the
circumference $2r$ goes to infinity and we get a two-dimensional
model defined on a plane. Since the gauge-fixed theory is defined
on a plane the asymptotic states and S-matrix are well-defined.
This limit has been studied in 
%\cite{AJK,Janik,Klose,AT,AF06,HM}.
\cite{AJK}-\cite{HM}.  An important observation recently made in
\cite{HM} is that in the limit one can give up the level-matching
condition and consider configurations with arbitrary world-sheet
momenta. Then, a one-soliton solution of the gauge-fixed string sigma
model should be identified with a one-magnon state in the spin chain
description of the gauge/string theory \cite{MZ,BDS,AFS,BS}, and the
world-sheet momentum is just equal to the momentum of the magnon \bea
p_{{\rm ws}} = p_{{\rm magnon}} = \D x_-\,.  \eea The corresponding
one-soliton solutions were named giant magnons in \cite{HM} because
generically their size is of order of the radius of ${\rm S}^5$. Since
for a giant magnon $\D x_-$ is not an integer multiple of $2\pi$, such
a soliton configuration does not describe a closed string.  It was
shown in \cite{HM} that the classical energy of a string giant magnon
is related to the momentum $p_{{\rm ws}}$ by the formula
\bea\la{HMdis} E_{{\rm string}}= {\sqrt{\lambda}\over
\pi}\bigg|\sin{p_{{\rm ws}}\ov 2}\bigg|\,, \eea which is the strong
coupling, (i.e.~$\l\to\infty$) limit of the spin chain dispersion
relation \cite{BDS} \bea\la{spindis} E_{{\rm spin\ chain}}=
\sqrt{1+{\lambda\over \pi^2}\sin^2{p\ov 2}}\,,\qquad p\equiv p_{{\rm
magnon}}\,.  \eea This is an interesting result because the appearance
of trigonometric functions is usually associated with a lattice
structure, while here the dispersion relation was derived in a
continuous model.  Moreover, the semi-classical S-matrix was also
computed in \cite{HM}, and shown to coincide with the semi-classical
approximation of the quantum string Bethe ansatz S-matrix of
\cite{AFS}.

\medskip

In this paper we want to stress  that it is natural to give up the
level-matching condition not only in the decompactifying limit but
also for finite $P_+$. The reason is that to quantize string theory in
a uniform gauge one has to consider all states with periodic $x^i$,
and impose the level-matching condition only at the end to single out
the physical subspace.  In a uniform gauge one still has a
well-defined model on a cylinder, however, if a string does not
satisfy the level-matching condition then its target space-time image
is an open string with end-points of the string moving in unison so
that $\D x_-$ remains constant. Another subtlety is that it is the
level-matching condition that makes gauge-fixed string sigma models
equivalent for different choices of a uniform gauge, that is for
different values of $a$. String configurations which do not satisfy
the level-matching condition may depend on $a$. This gauge-dependence
makes the problem of quantizing string theory in a uniform gauge very
subtle. On the other hand the requirement that physical states are
gauge independent should impose severe constraints on the structure of
the theory. It may also happen that for finite $J$ there is a
preferred choice of the parameter $a$ simplifying the exact
quantization of the model. In fact we will see that for finite $J$ one
can identify the world-sheet momentum (\ref{pws}) with a spin-chain
magnon momentum only in the $a=0$ gauge. This seems to make $a=0$ gauge
choice the most natural one at least in the AdS/CFT context. Furthermore,
this gauge is also distinguished because only in the uniform gauge
 one can study string configurations with an arbitrary
winding number in one go. In this respect it is closer to the
conformal gauge, and we will see that the one-magnon energy is in fact
the same in these two gauges.

In the strong coupling limit $\l\to\infty$ one should be able to use
the classical string theory to find a corresponding one-soliton
solution\footnote{Strictly speaking, since for finite $P_+$ the theory
is defined on a cylinder, the corresponding solution should be
probably viewed as a superposition of an infinite number of usual
solitons on a plane. We will see, however, that these one-soliton
solutions are uniquely determined even for finite $P_+$.} and
determine the finite $P_+$ corrections to the dispersion relation
(\ref{HMdis}). This is the problem we are going to address in the next
sections.

%%%%%%%%%%%%%%%%%%%%%%%%%%%%%%%%%%%%%%%%%%%%%%%%%%%%%%%%%%%%%%
\section{Giant magnon in uniform gauge}
%%%%%%%%%%%%%%%%%%%%%%%%%%%%%%%%%%%%%%%%%%%%%%%%%%%%%%%%%%%%%%
As was discussed in \cite{HM}, a giant magnon is a string moving
on a two-dimensional sphere. This is a consistent reduction
of classical string theory on $\AdS$.
Our starting point is the bosonic
action (\ref{S1}) for strings in ${\mathbb R}\times {\rm S}^2$
\bea\la{ss2} S= -{\sqrt\l\ov 4\pi}\int_{-r}^{ r}\, {\rm d}\s{\rm
d}\tau\, \g^{\a\b}\left(-\pa_\a t\pa_\b t + \pa_\a X_i\pa_\b X_i
\right)\,, \eea where $X_i X_i = 1$. We find convenient to use the
following parametrization of ${\rm S}^2$ \bea X_1 + i X_2 =
\sqrt{1-z^2} e^{i\p}\,,\quad X_3 = z\,,\quad -1\le z\le 1\,. \eea
The coordinate $z$ is related to the standard angle $\theta$ as $z
= \cos \theta$. The values $z=\pm 1$ correspond to the north and
south poles of the sphere,  and at $z=0$ the angle $\phi$
parametrizes the equator. In terms of the coordinates $\p$ and $z$
the metric of $S^2$ takes the form \bea ds^2_{S^2} = {dz^2\ov
1-z^2} + (1-z^2) d\p^2\,. \eea

%%%%%%%%%%%%%%%%%%%%%%%%%%%%%%%%%%%%%%%%%%%%%%%%%
\subsection{Soliton solution}
%%%%%%%%%%%%%%%%%%%%%%%%%%%%%%%%%%%%%%%%%%%%%%

Introducing the light-cone coordinates (\ref{lcc}), imposing the uniform
gauge (\ref{ulc}) (with $m=0$), and following the steps described
in the previous section, we derive the gauge-fixed string action
\bea \la{Su1} S ={\sqrt{\lambda}\over 2\pi} \int_{- r}^{ r}\, {\rm
d}\s{\rm d}\tau\, \left( p_z \dot{z} \,-\, \H \right)\, , \eea
where the density of the gauge-fixed Hamiltonian is a function of
the coordinate $z$ and its canonically conjugate momentum $p_z$. 
Recall also that $r =  {\pi\ov\sqrt{\lambda}}P_+ =
{\pi\ov\sqrt{\lambda}}\left((1-a)J + aE\right)$. Explicit
expressions for the Hamiltonian and other quantities computed in
this section can be found in Appendix A where we also present
their forms for the three simplest cases $a=0,1/2,1$.

\medskip

To find a one-soliton solution of the gauge-fixed string theory it is convenient to
go to the Lagrangian description by eliminating the momentum $p_z$.
Solving the equation of motion for $p_z$ that follows from the action (\ref{Su1}), we determine the momentum as a function of $\dz$ and $z$. Then
substituting the solution into (\ref{Su1}), we obtain the action in the Lagrangian form: $S=S(z,z',\dz)$.
The explicit form of the action is given in Appendix A, and it is of the Nambu-Goto form. We will see in a moment that this leads to the existence of
finite-energy singular solitons.

To find a one-soliton solution we make the most general ansatz
describing a wave propagating along the string \bea \la{ansatz} z =
z(\s - v\tau)\,, \eea where $v$ is the velocity of the soliton.
Substituting the ansatz into the action (\ref{Su2}), we derive the
Lagrangian, $L_{red}=L_{red}(z,z')$, of a reduced model which defines
a one-particle system if we regard $\s$ as a time variable.  The
$\s$-evolution of this system can be easily determined by introducing
the ``momentum'' conjugated to $z$ with respect to ``time'' $\s$
\bea\nonumber \pi_z = {\pa L_{red}\ov \pa z'} \,, \eea and computing
the reduced Hamiltonian \bea\nonumber H_{red} =\pi_z z' - L_{red} \,.
\eea The reduced Hamiltonian is a conserved quantity with respect to
``times'' $\s$, and we set it to some constant \bea\nonumber H_{red} =
{\om -1\ov 1 - a + a\, \om}\, .  \eea Here we have chosen to
parametrise the constant $H_{red}$ in this way in order to simplify
the comparison with the conformal gauge solution in section \ref{cg}.

Solving this equation with respect to $z'$, we get the following basic
equation
\bea\la{zp}
z'^2 = \left(\frac{1-z^2}{(1-a)
\left(b^2-z^2\right)}\right)^2\, \frac{ z^2-z_{min}^2}{z_{max}^2-z^2} \ ,
\eea
where the parameters $z_{min}$, $z_{max}$ and $b$ are related to $a$, $v$ and $\om$ as follows
\bea
z_{min}^2 = 1-{1\ov \om^2}\,,\qquad z_{max}^2=1-v^2\,,\qquad
b^2 = 1+{a\ov (1-a)\, \om}\,.
\eea
A one-soliton solution we are looking for corresponds to a periodic solution of the equation
(\ref{zp}),
the period being equal to $2 r = {2\pi\ov\sqrt{\lambda}}P_+$.
The parameter $z_{min}$ is determined by the period of the solution.

\medskip

It is not difficult to see that such a solution exists if
the following inequalities hold
\bea\la{rang}
0\le a\le 1\,,\quad 0\le z_{min}^2\le z_{max}^2\le
b^2\,.
\eea
It follows from these inequalities that the range of $a,\om$ and $v$ is
\bea\la{rangomv}
0\le a\le 1\,,\quad 1\le \om < \infty\,,\quad    0\le |v|\le {1\ov \om}\,.
\eea

Then, assuming for definiteness that $z\ge 0$,
the corresponding solution of the  equation (\ref{zp}) lies between $z_{min}$ and $z_{max}$,
and for given $a$ and $v$ the parameter $z_{min}$ is found from the equation
\bea\la{period}
r = \int_0^r {\rm d}\s = \int_{z_{min}}^{z_{max}} {{\rm d}z\ov |z'|}\,.
\eea
This integral can be easily computed in terms of elliptic functions by using formulas from Appendix B.

\medskip

One can easily see from equation (\ref{zp}) that in the range of parameters (\ref{rang}) the shape of the soliton is similar for any values of $a,v$ and $\om$. The allowed values of $z$ are
$z_{min}\le z\le z_{max}$, and $z'$ vanishes at $z=z_{min}$, and goes to infinity
at $z=z_{max}$. So, if we assume that at $\tau = 0$ the solution is such that
$z'=0$ at $\s = -r$ and $\s=r$, then
$z = z_{max}$ at $\s =0$, and the soliton profile is shown in Fig.(\ref{zprofile}).

\begin{figure}[t]
\begin{center}
\psfrag{z}{$z$}
\psfrag{sigma}{$\sigma$}
\includegraphics*[width=.45\textwidth]{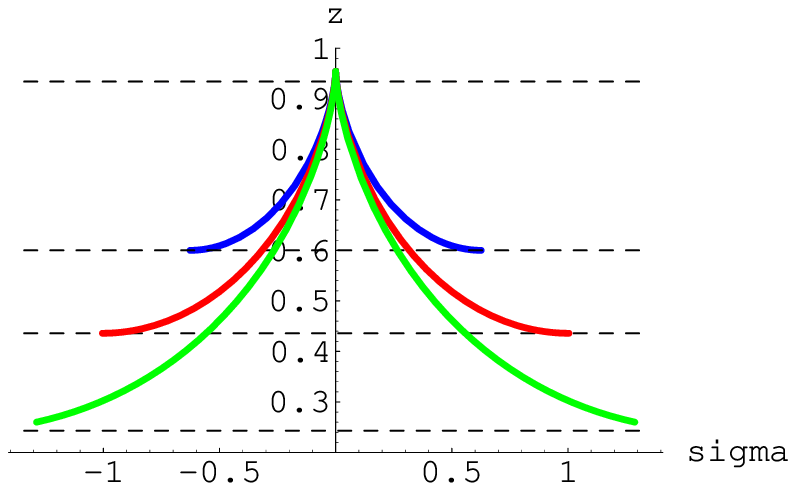}
\psfrag{xm}{$x^-$}
\psfrag{sigma}{$\sigma$}
\includegraphics*[width=.45\textwidth]{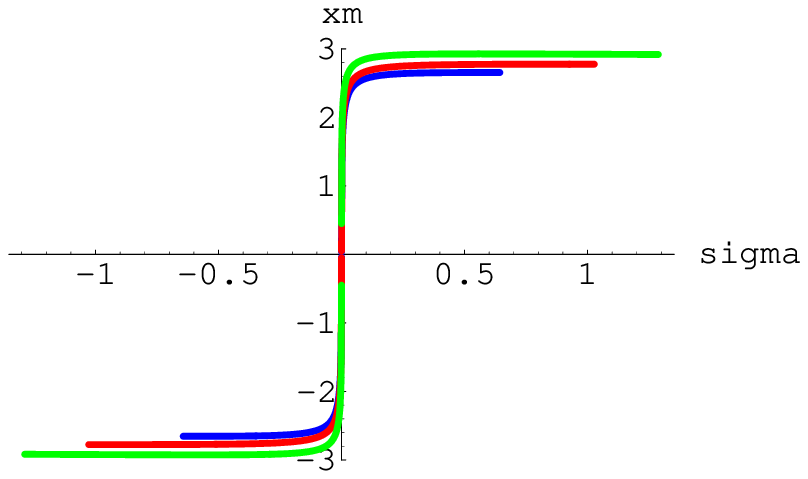}
\end{center}
\caption{\la{zprofile}Profile of $a=0$ one-magnon soliton: Left, $z(\sigma)$ plotted for configurations with the same $z_{\text{max}}=0.99$  and $z_{\text{min}}= \{0.6,0.19,0.06 \}$, green, red and blue respectively.  Right, profile $x^-(\sigma)$ for the same values of  
$z_{\text{min}}, z_{\text{max}}$.}
\end{figure}

\begin{figure}[t]
\begin{center}
\psfrag{z}{$z$}
\psfrag{xm}{$x^-$}
\includegraphics*[width=.7\textwidth]{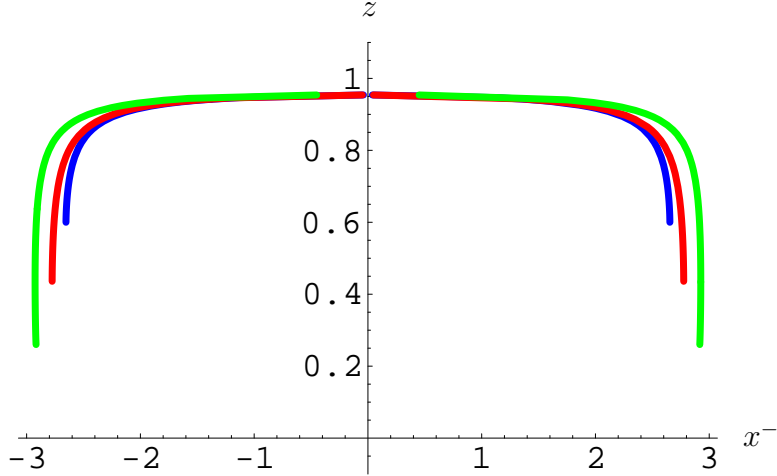}
\end{center}
\caption{\la{shape-a=0}Target space shape of magnon at fixed light-cone time $x^{+}$, depicted for three magnons moving in the stripe $z_{\text{max}} = 0.99$, $z_{\text{min}} = \{0.6,0.19,0.06 \}$, green, red and blue respectively.} 
\end{figure}
The corresponding solution is, as we see, not smooth at $z= z_{max}$. The energy of
this soliton is however  finite. To compute the energy, we need to evaluate $\H/|z'|$ on the solution:
\bea\nonumber
{\H\ov |z'|} = {
z^2 -(\om-1)\left( {1\ov \om} +{(1-a)v^2 \ov 1-a + a \om}\right) -
{v^2\ov 1 - z^2}{a(\om-1)\ov \om(1-a+ a \om)}\ov\sqrt{(z_{max}^2 - z^2)(z^2 -z_{min}^2) }}\,.
\eea
Then the energy of the soliton is given by the following integral
\bea\la{energya}
E-J = {\sqrt{\lambda}\over 2\pi}\int_{- r}^{ r}\, {\rm d}\s\, \H =
{\sqrt{\lambda}\over \pi}\int_{z_{min}}^{z_{max}}\, {\rm d}z {\H\ov |z'|}\,,
\eea
and it is clear from this expression that the energy is finite.

Finally, we also need to compute the world-sheet momentum (\ref{pws})
\bea\la{pwsu}
\pws = -\int_{-r}^{r} d\s p_z z' = 2\int_{z_{min}}^{z_{max}}\, {\rm d}z |p_z|\,,
\eea
where we have assumed that $v>0$, and took into account that then
for the soliton we consider the product
$-p_z z'$ is positive. The following explicit formula for the momentum $p_z$
canonically conjugate to $z$
can be easily found by using equation.(\ref{pza}),(\ref{ansatz}) and (\ref{zp})
\bea\la{pzu}
p_z ={v\om\ov 1-a+ a \om}\, \frac{1}{1-z^2}\, \frac{\sqrt{z^2 - z_{min}^2}}{\sqrt{z_{max}^2-z^2}}\,.
\eea
Let us also mention that in the case of a one-soliton solution the world-sheet momentum (\ref{pwsu})
is just equal to the canonical momentum carried by the center of mass of the soliton. To see that we
just need to plug the ansatz (\ref{ansatz}) into the string action (\ref{Su1}), and integrate over $\s$.
Then we obtain the following action for a point particle
$$
S = {\sqrt{\lambda}\over 2\pi}
\int {\rm d}\tau\, \left(
\pws\, v \,-\, \H \right)\,,
$$
that shows explicitly that $\pws$ is the soliton momentum.

It is also useful to understand the target-space shape of the soliton,
that is to find the dependence of $z$ on the target-space coordinate
$x_-$. To this end we compute the derivative $dz/dx_-$ \bea\la{dzdxm}
\bigg|{dz\ov dx_-}\bigg|= \bigg|{z'\ov x_-'}\bigg| = {1\ov p_z} =
\frac{1-a+ a \om}{v\om}\, (1-z^2)\frac{\sqrt{z_{max}^2-z^2}}{\sqrt{z^2
- z_{min}^2}}\,.  \eea We see that in the target space, the soliton
configuration is in fact smooth at $z=z_{max}$ and singular at
$z=z_{min}$. Then the configuration is not static, see
Fig.\ref{zevolv} and the simulations
at~\url{http://www.aei.mpg.de/~peekas/magnons/}.
\begin{figure}[t]
\begin{center}
\psfrag{Magnon}{\smaller Magnon}
\psfrag{Path}{\smaller Path}
\psfrag{Equator}{\smaller Equator}
\psfrag{1/om1}{\smaller\smaller $\sin \theta = \frac{1}{\omega}$}
\psfrag{v/om1}{\smaller\smaller $\sin \theta = v$}
\psfrag{Dphi}{$\Delta \varphi$}
\includegraphics*[width=.45\textwidth]{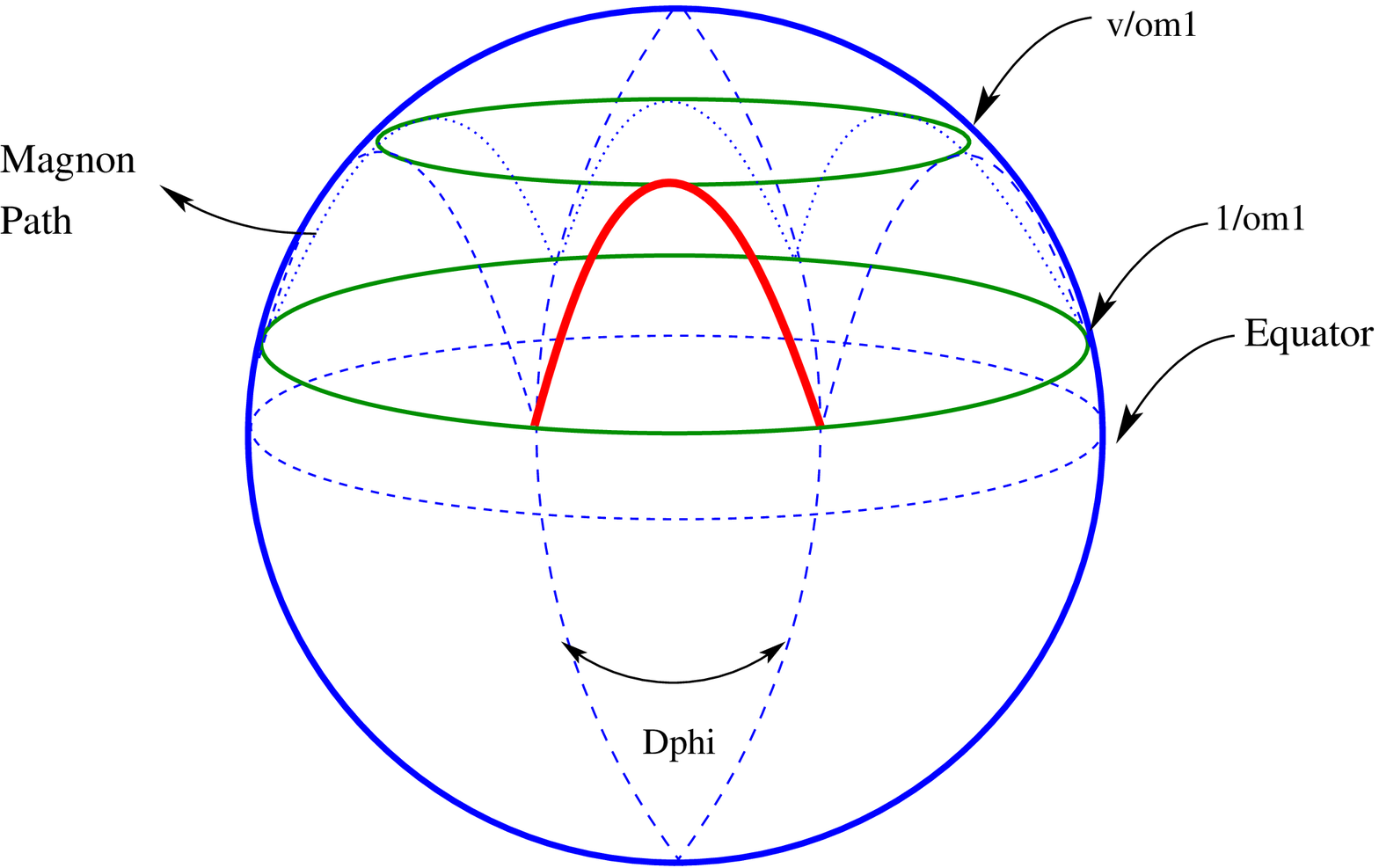}
\qquad
\psfrag{1/om}{\smaller\smaller $\sin \theta = \frac{1}{\omega}$}
\psfrag{v/om}{\smaller \smaller$\sin \theta = v$}
\includegraphics*[width=.45\textwidth]{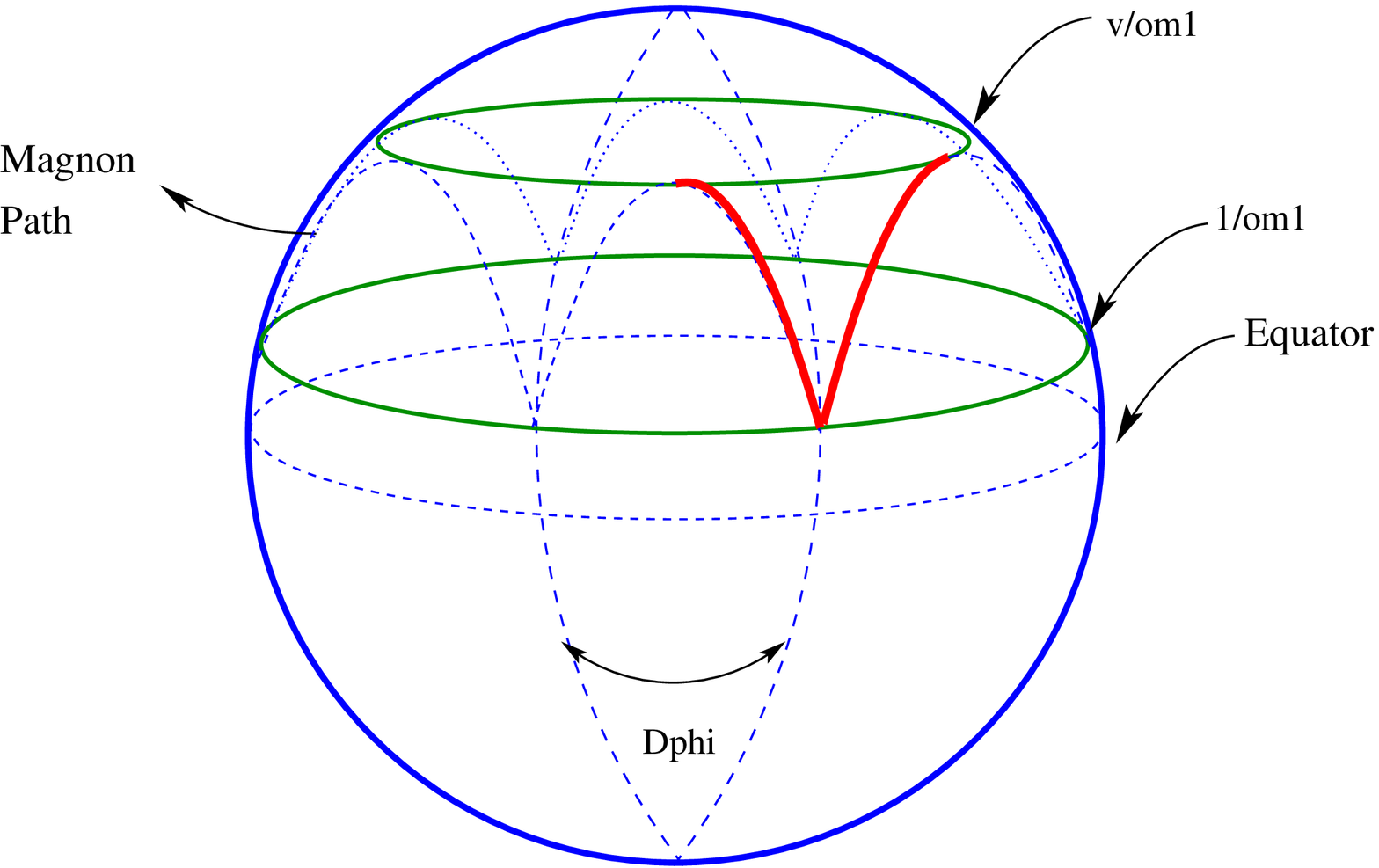}
\end{center}
\caption{\la{zevolv}Plot of the time evolution of the finite $J$ magnon on the sphere. Left picture depicts moment $t=0$, while the right $t=T/4$ , where $T$ is period: $T=2r/v$. We see that string develops a spike during the time evolution.}
\end{figure}
%%%%%%%%%%%%%%%%%%%%%%%%%%%%%%%%%%%%%%%%%%%%%%%%%
\subsection{Infinite $J$ giant magnon}
%%%%%%%%%%%%%%%%%%%%%%%%%%%%%%%%%%%%%%%%%%%%%%

To recover the uniform gauge equivalent of the giant magnon solution of \cite{HM} we need
to send $r$ to infinity. It is easy to see that this limit corresponds to setting the
parameter $z_{min}$ to 0 or equivalently $\om$ to 1.
In the limit $\om\to 1$ the basic equation (\ref{zp}) simplifies
\bea\la{zp0}
z'^2 = \left(\frac{z(1-z^2)}{\left(1-(1-a)z^2\right)}\right)^2\,
\frac{1}{1-v^2-z^2} \ ,
\eea
and can be easily integrated. The range of $\s$ is now from $-\infty$ to $\infty$, and
both $z'$ and $z$ vanish at $\s=\pm\infty$.  Even though the solution $z(\s -v\tau)$ depends on
$a$, for all values of $a$ from the interval $[0,1]$ the energy (\ref{energya})
and the world-sheet momentum (\ref{pwsu}) depend only on $v$
\bea\la{energyaQ0}\nonumber
E-J &=&
{\sqrt{\lambda}\over \pi}\int_{z_{min}}^{z_{max}}\, {\rm d}z {\H\ov |z'|}
= {\sqrt{\lambda}\over \pi}\int_{0}^{\sqrt{1-v^2}}\, {\rm d}z {z\ov \sqrt{1-v^2-z^2}} =
{\sqrt{\lambda}\over \pi}\sqrt{1-v^2}\,,\\
\la{pQ0}\nonumber
\pws &=& 2\int_{z_{min}}^{z_{max}}\, {\rm d}z |p_z|=2\int_{0}^{\sqrt{1-v^2}}\, {\rm d}z
\frac{v z}{ \left(1-z^2\right)\sqrt{1-v^2-z^2}}=2\arccos v\,.
\eea
Expressing $v$ in terms of $\pws$, we get the dispersion relation (\ref{HMdis}). This demonstrates explicitly that in the infinite $J$ limit one can give up the level-matching condition and still have
independence of the gauge choice. We will see in the next subsection that it is not the case for
finite $J$.

%%%%%%%%%%%%%%%%%%%%%%%%%%%%%%%%%%%%%%%%%%%%%%%%%
\subsection{Finite $J$ giant magnon}
%%%%%%%%%%%%%%%%%%%%%%%%%%%%%%%%%%%%%%%%%%%%%%

To find the dispersion relation for finite $J$ we need to express the
soliton energy $E-J$ in terms of $J$ and the world-sheet momentum
$\pws$. It is obvious that there is no simple analytic expression for
the dispersion relation. It is possible however to analyze dispersion
relation for large values of the charge $J$. The details of this
complicated analysis are given in Appendix B. All corrections turn out
to be only exponential in this limit. The  leading and  subleading exponential corrections to $E-J$ computed in the appendix are
\bea\label{ge3} E-J&=&\frac{\sqrt{\lambda}}{\pi}\sin\frac{\pws}{2}
\Big[1-\frac{4}{e^2}\sin^2\frac{\pws}{2}~ e^{-\j}-\\ \nonumber
&&~~~~~~-\frac{4}{e^4}\sin^2\frac{\pws}{2}~\Big( \j^2(1+\cos
\pws)+2\j(2+3\cos \pws+a\pws\sin \pws) +\\
&&~~~~~~~~+7+6\cos \pws +6 a\pws\sin \pws+a^2\pws^2(1-\cos \pws)
\Big)e^{-2\j}+\cdots \Big]\, . \nonumber \eea
Here we have introduced
the effective length felt by the magnon with momentum $\pws$
\bea\la{calJ}
\j =\frac{2\pi J}{\sqrt\l\sin\frac{\pws}{2}} +a\pws\cot\frac{\pws}{2}\,.
\eea
Formula (\ref{ge3}) has several interesting features. First of all it shows that
the exponential correction is basically determined by the ratio $J/(E-J)$ because
for large values of $J$, $\j \sim 2J/(E-J)$. Then,
for generic values of $a$
the dispersion relation is not periodic in $\pws$. The periodicity in $\pws$ is restored only
for $a=0$. This indicates that for finite $J$
one can identify the world-sheet momentum with a spin-chain magnon momentum only for $a=0$.

Formula (\ref{ge3}) also shows a nontrivial dependence on the
parameter $a$. Only the coefficients of the leading terms, $e^{-\j}$ and $\j^2
e^{-2\j}$, are independent of $a$. The dependence must, however,
disappear in the case  $v = 0$ which corresponds to the finite-$J$
generalization of the ``half-GKP" solution \cite{GKP} describing
an open string satisfying the Neumann boundary conditions and 
rotating on ${\rm S}^2$ with spin $J$. 
%At infinite $J$ this solution corresponds to $\pws=\pi$. 
Computing the world-sheet momentum in the
limit $v\to 0$, we find
\bea
\pws\to  {\pi\ov 1-a + a\om}\, .
\label{pja3}
\eea
Thus, $\pws\to \pi$ only in the case $a=0$ or in the case $\om=1$ that corresponds
to the infinite $J$ limit.
Remembering that the momentum $p_{{\rm magnon}}$ 
of a spin-chain magnon changes from $-\pi$ to $\pi$ (we consider a zero-winding case), we see 
that $\pws$ can be naturally identified with $p_{{\rm magnon}}$ only for $a=0$.\footnote{In principle one could rescale the momentum by a factor depending on $J$ and $a$ so that the rescaled momentum would have the same range for any $a$. This rescaling, however, would mean a rescaling of the world-sheet coordinates $\tau$ and $\s$, and,
as a result, the world-sheet Hamiltonian would not be equal to $E-J$. }
Now taking into account that $\om$ is a function of $a$ and $J$, we find that in the limit
$v\to 0$  the world-sheet momentum has the following expansion
\bea
\pws=\pi-\frac{8\pi a}{e^2}e^{-\cj}+\frac{32\pi
a(-1+2a+\cj)}{e^4}e^{-2\cj}+\cdots\, , \label{expansionp3}
\eea
Substituting this formula into (\ref{ge3}), we obtain 
the following formula for the exponential correction to
the energy of the ``half-GKP" solution 
\bea \label{halfGKPen3}
E-J=\frac{\sqrt{\lambda}}{\pi}\Big[1-\frac{4}{e^2}
e^{-\cj}-\frac{4}{e^4}(1-2\cj) e^{-2\cj}+\cdots \Big]\, , 
\eea
where $\cj=\frac{2\pi J}{\sqrt\l}$.
This expression has no $a$-dependence  as it should be. 
The fact that there is no $a$-dependence in this case also follows from 
exact equations (\ref{relf}) and (\ref{energie}) in Appendix B without 
performing any expansion.

Let us also mention that the GKP folded string rotating on ${\rm
S}^2$ with spin $J$ \cite{GKP} can be thought of as being composed
from two giant magnons with spin $J/2$ and $\pws=\pi$. The energy
of the folded string is still equal to the sum of energies of the
magnons even at finite $J$. In fact, in the $a=0$ gauge if $\pws =
{2\pi m\ov N}$ we can also build a closed string configuration
with the winding number $m$ carrying the charge $J$ by gluing $N$
finite $J/N$ solitons, see section \ref{cg} for a discussion of
this configuration, and Fig.(\ref{star1}), (\ref{star}) and (\ref{Dphi-plot2}).
The resulting configuration was analyzed in \cite{Ryang}, and is an ${\rm S}^2$-analogue of the
$\rm{AdS}_3$ spiky strings studied in \cite{Krspiky}, and the
energy of this closed string is again equal to the sum of energies
of the $N$ magnons.

In general, however, we expect the simple addition formula for the
energy of a composite closed string to be correct only at infinite $J$
where all the exponential corrections in (\ref{ge3}) vanish.  If so
then at finite $J$ the string spectrum would not be described by a
simple Bethe ansatz of the form \cite{AFS}. If a Bethe ansatz
description of the string spectrum is possible at all then it would
have auxiliary excitations and a more complicated dispersion relation
with the usual one (\ref{spindis}) arising only at infinite $J$
similar to what happens in the Hubbard model description of the BDS
spin chain \cite{RSS}.  It is worth noting that in the Hubbard model
the exponential corrections to the dispersion relation are also
governed by the same effective length $\j$ (\ref{calJ}) (with
$a=0$). However, one can check, that these corrections appear only in
the subleading orders in $1/\sqrt\l$. The strong coupling dispersion
relation (\ref{HMdis}) is not modified in the Hubbard model, unlike
what is observed here. Similarly, the \emph{quantum}, finite size
corrections computed in \cite{AJK}, seems to predict that the same
exponential term governs finite size corrections at the quantum level.

It is also worth stressing that at large $\l$ a realistic quantum
string Bethe ansatz should lead to the same exponential correction for
the dispersion relation, and that our result should serve as a
nontrivial check of any proposal for such an ansatz. Note also that not
only exponential terms, but also highly non-trivial coefficients
multiplying series in $\j$ need to be reproduced.

%%%%%%%%%%%%%%%%%%%%%%%%%%%%%%%%%%%%%%%%%%%%%%%%%%
\section{Global symmetry algebra}\la{ch}
%%%%%%%%%%%%%%%%%%%%%%%%%%%%%%%%%%%%%%%%%%%%%%%%%%
In this section we discuss the implications of giving up the level-matching condition for
the global symmetry algebra of a string model.

Recall that the theory we consider is obtained by reduction of the
string sigma-model on $\AdS$ to a smaller space $\mathbb{R}\times
{\rm S}^2$. This space still has a non-trivial isometry group
which is $\mathbb{R}\times {\rm SO}(3)$, where $\mathbb{R}$
corresponds to the shifts of the global AdS time $t$ and ${\rm
SO}(3)$ is the isometry group of the two-sphere. It is known
that giving up the level-matching condition leads to dramatic
consequences for the global symmetry algebra, namely, it gets
reduced, because some of the global charges fail to satisfy the
conservation law.

\bigskip

This phenomenon is of course general and it also occurs for closed
strings propagating in flat space. Indeed, in the light-cone gauge
the dynamical generators of the Lorentz algebra are given by
$$
J^{i-} =\int_0^{2\pi} {\rm d}\sigma \, (X^i \dot{X}^- - X^-
\dot{X}^i) \, .
$$
Using the flat string equations of motion $\Box X^M=0$ for $M=i,-$
the (total) time derivative of these generators can be reduced to
the total derivative term
\begin{eqnarray}
\label{dJ} \dot{J}^{i-} &=&  \int_{0}^{2\pi} {\rm d}\sigma\, ( X^i
\ddot{X}^- - \ddot{X}^i X^-)
            =   - {X^i}'(0) \Big(X^-(2\pi) - X^-(0)\Big)  \nonumber \, ,
\end{eqnarray}
where we have used the fact that the transversal fields $X^i,
(X^i)'$ are periodic. If the level-matching condition is not
satisfied, i.e. $\Delta X^-=X^-(2\pi) - X^-(0)\neq 0$ the
dynamical generators of the Lorentz algebra are broken. Only for
special configurations, for which the transversal coordinates obey
the {\it open} string condition ${X^i}'(0)=0={X^i}'(2\pi)$ the
dynamical generators in question are still conserved. This picture
has a clear physical meaning: As soon as we give up the
level-matching condition, the coordinate $X^-$ becomes distinguished from
the periodic transversal coordinates $X^i$ and this leads to
non-conservation of the Lorentz algebra generators which mix $X^-$
with transversal directions.

\bigskip

This discussion can be easily generalized to the uniform gauge for
strings on ${\mathbb R}\times {\rm S}^2$. The Noether charges of
the global ${\rm SO}(3)$ symmetry are defined as
$$
J_{MN}=\frac{\sqrt{\lambda}}{2\pi}\int_{-r}^r \frac{{\rm
d}\sigma}{2\pi} \gamma^{\tau\a}\pa_{\a}X_{[M} X_{N]}\, ,
$$
where $M,N=1,2,3$ and $X^M$ are defined as (cf. section 3)
$$
X_1=\sqrt{1-z^2}\cos\phi\, , ~~~~~~X_2=\sqrt{1-z^2}\sin\phi\, ,
~~~~~~~X_3=z\, .
$$
The time-derivative of the charges can be again written as the
total derivative by using equations of motion for the fields $X^M$
and we get \bea
\dot{J}_{MN}=-\frac{\sqrt{\lambda}}{2\pi}\int_{-r}^r {\rm d}\sigma
\pa_{\sigma}\Big( \gamma^{\sigma\a}\pa_{\a}X_{[M}
X_{N]}\Big)=-\frac{\sqrt{\lambda}}{2\pi}\Big(\gamma^{\sigma\a}\pa_{\a}X_{[M}
X_{N]}\Big)\bigg |_{\sigma=-r}^{\sigma=r}\, . \label{nonc} \eea
The components of the worlds-sheet metric can be found from the
action (\ref{S3}) by considering equations of motion for $p_{\pm}$
and $x_-$, see \cite{AF} for a detail discussion in the $a=0$ case
 \bea
 \nonumber
\gamma^{\tau\tau}&=&a\frac{a{\cal H}-1}{1-z^2}-(1-a)(1+(1-a){\cal
H})\, ,\\
\nonumber \gamma^{\tau\sigma}&=&p_zz'(1-2a-(1-a)^2 z^2)\, .
 \eea
We see, in particular, that the metric components do not involve
the unphysical field $x_-$ and that for our soliton solution they
are periodic functions of $\sigma$. The r.h.s. of equation (\ref{nonc})
contains also the $\tau$- and $\sigma$-derivatives of the field
$x_-$ which are \bea \nonumber
 \dot{x}_-&=&\frac{-{\cal H}+(1+{\cal H}-a{\cal
H})z^2-p_z^2z'^2(1-z^2)(1-2a-(1-a)^2z^2)}{1+{\cal H}-2a{\cal
H}-(1-a)(1+{\cal H}-a{\cal H})z^2}\, , \\
\nonumber x'_-&=&-p_zz'\, .
 \eea
Again, the r.h.s. of these equation do not involve $x^-$ itself,
the field responsible for the violation of the level-matching
condition. Plugging everything into equation (\ref{nonc}) we first
verify that the generator $J_{12}=J$ is conserved. This is
the generator corresponding to the isometry $\phi\to\phi+{\rm
const}$. However, the time derivatives of the (non-diagonal)
generators $J_{13}$ and $J_{23}$ involve $\sin\phi$ and
$\cos\phi$, and, since
$$
\phi=\tau+(1-a)x_-\,
$$
they are not periodic functions of $\sigma$ because $x^-$ is not
periodic; $\Delta\phi=(1-a)\Delta x_-=(1-a)p_{\rm ws}$ for our
soliton solution. Note, however, that all these charges are conserved in the
$a=1$ case where $\phi=\tau$.

One can further see that the expression for the
time derivatives of the non-diagonal generators is proportional to
$z'(\sigma-v\tau)$ which should be evaluated at $\sigma=\pm r$. In
the infinite $J$ case, when $r\to \infty$, this
derivative vanishes for any finite $\tau$ and this leads to
conservation of all the charges for a giant magnon. In other words,
an infinite $J$ giant magnon satisfies open string boundary conditions which
allow for the unbroken symmetry algebra. In our case of finite
$r$, the derivative of $z$ vanishes only at $\tau=0$ and,
therefore, the non-diagonal symmetry generators are broken.

%%%%%%%%%%%%%%%%%%%%%%%%%%%%%%%%%%%%%%%%%%%%%%%%%%%%%%%%%%%%%%
\section{Giant magnon in the conformal gauge}\la{cg}
%%%%%%%%%%%%%%%%%%%%%%%%%%%%%%%%%%%%%%%%%%%%%%%%%%%%%%%%%%%%%%
In this section we discuss the finite $J$ giant magnon in the
conformal gauge generalizing the consideration of \cite{HM}. 
It is well-known that string theory on ${\mathbb R}\times {\rm S}^2$ 
in the conformal gauge can be mapped to the sine-Gordon model 
\cite{Pohl, Mik},\footnote{ 
The map, however, does not preserve the Poisson structure \cite{Mik2}, 
and, by this reason, the 
two models describe different physics. }
that can be used to find multi-soliton solutions in the string theory. We find it, however, 
simpler to obtain the giant magnon solutions directly from string theory on ${\mathbb R}\times {\rm S}^2$.

We
start with the same action (\ref{ss2}) for strings in ${\mathbb
R}\times {\rm S}^2$, and impose the conformal gauge $\g_{\mu\nu} =
{\mbox {diag}}(-1,1)$, and the condition $t =\tau$. Then the
world-sheet space coordinate $\s$ must have the range \bea
\label{where} -r\le \s\le r\,, \qquad r = {\pi\ov\sqrt\l}E\,, \eea
where $E$ is the target space-time energy. Note that this is the
same range as in the $a=1$ uniform gauge. The condition $t=\tau$,
however, corresponds to the $a=0$ gauge.

The gauge-fixed action takes the form
\bea\la{Scg}
S = -{\sqrt{\lambda}\over 4\pi}\int_{-r}^{ r}\, {\rm d}\s{\rm d}\tau\,
 \left({\pa_\mu z \pa_\mu z\ov 1-z^2} + (1-z^2)\pa_\mu\p \pa_\mu\p\right)\,,
\eea
and the equations of motion that follow from the action should be supplemented by
the Virasoro constraints
\bea\la{Vc}
&&{\dz^2+z'^2\ov 1-z^2} +(1-z^2)\left( \dph^2 + \p'^2\right) = 1\,,\\
&& {\dz z'\ov 1-z^2} +(1-z^2)\dph\p'= 0\,.
\eea
The invariance of the action under shifts of the angle $\p$ leads to the existence of the conserved
charge $J$
\bea\la{Jcg}
J = {\sqrt{\lambda}\over 2\pi}\int_{-r}^{ r}\, {\rm d}\s\, (1-z^2) \dph \,.
\eea

\begin{figure}[t]
\begin{center}
\psfrag{z}{z}
\psfrag{t}{t}
\includegraphics*[width=.44\textwidth]{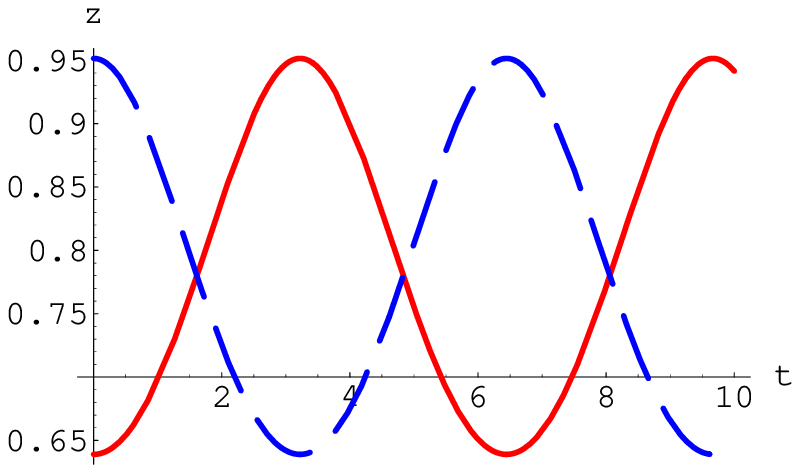}
\includegraphics*[width=.44\textwidth]{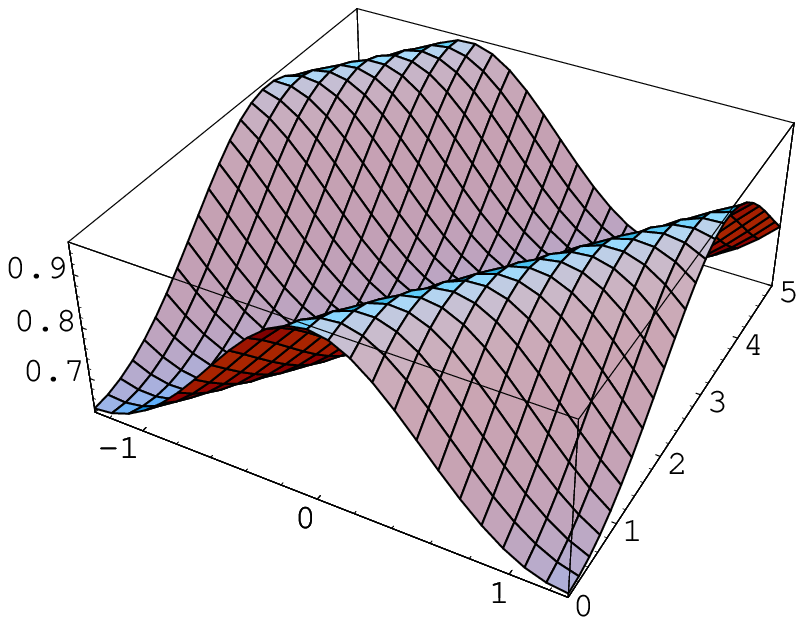}
\end{center}
\caption{\label{ez}First picture: plot of the time evolution of the end point and middle point of
finite $J$ magnon in the $z$ direction. 
The red (solid) line are string end points, while the blue (dashed) line is the middle point of the string. Plot is made for $\omega = 1.3$ and $v=0.4$
Second picture: plot of time evolution of string in the $z$ direction; 
axis $x = \sigma$, $z=\cos\theta$, $y$-time.
}
\end{figure}

\begin{figure}[t]
\begin{center}
\includegraphics*[width=.6\textwidth]{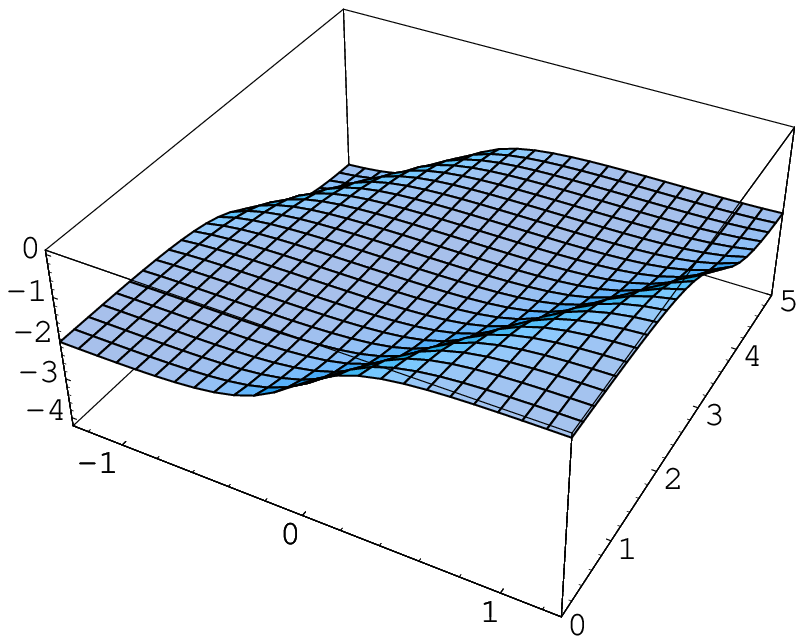}
\end{center}
\caption{\label{ep}Plot of time evolution of string in $\phi$ direction; axis are labeled as: $x = \sigma$, 
$z= \phi$, $y$-time (parameters are $\omega = 1.3$ and $v=0.4$).
}
\end{figure}
\medskip
We will be looking for a solution of the equations of motion
satisfying the following boundary conditions \bea\la{bc} z(r,\tau)-
z(-r,\tau)=0\,,\quad \D\p=\p(r,\tau)- \p(-r,\tau)= p\,, \eea where $p$
is a constant which is identified with the magnon momentum \cite{HM}.
Since the field $\p$ does not satisfy the periodic boundary conditions
such a solution describes an open string.\footnote{Let us note that
the nonperiodicity of $\p$ (up to a winding number) is the reason why the giant magnon
solution cannot be found by using the KMMZ equations \cite{KMMZ}.} For
finite $E$ a justification to such a choice of boundary conditions
comes from the consideration of the string theory in the uniform gauge
$t=\tau$ where the world-sheet momentum (\ref{pws}) is equal to the
change of $\D x_-=\D\p$. One can see that these boundary conditions
are compatible with the equations of motion and Virasoro constraints.

The finite $E$ solution can be easily found by introducing a ``light-cone'' coordinate~$\vp$
\bea
\vp = \p  - \om t\,,
\eea
and taking the same one-soliton ansatz which was used in section 2
\bea
\vp = \vp(\s -v\om \tau)\,,\quad z = z(\s-v\om \tau)\,.
\eea
Note that the velocity of the soliton in the conformal gauge is
$v_{cg}=v \om$. We use this parametrization because, as we will see in
a moment, the parameters $\om$ and $v$ coincide with the corresponding
parameters in the $a=0$ uniform gauge. Recall that the parameter $\om$
should be greater than 1, and go to 1 as $E$ approaches infinity, as
can be also seen by analyzing the folded string solution of
\cite{GKP}, and $v^2 < 1/\om^2$.

For our ansatz the Virasoro constraints give the following equations
\bea
\label{dphi}
&&\vp' = {v\om^2\ov 1-\om^2 v^2} { z^2-z_{min}^2\ov 1-z^2}\,,\\
\label{dth}
&&z'^2 ={\om^2\ov (1-\om^2v^2)^2}\left(z^2-z_{min}^2\right) \left(z_{max}^2-z^2\right)\,,
\eea
where
\bea
z_{min}^2 = 1-{1\ov\om^2}\,,\qquad  z_{max}^2 = 1-v^2\,.
\eea
We see that for this solution the derivative $z'$ is finite everywhere,
and vanishes  both  for $z=z_{max}$ and $z=z_{min}$. This derivative, however, does not
have a gauge-invariant meaning. The real target-space shape of the solution is determined by
$dz/d\vp$, which vanishes at $z=z_{max}$ but diverges at $z=z_{min}$. The derivative is in fact
\emph{equal} to the derivative $dz/dx_-$ (\ref{dzdxm}) in the $a=0$ uniform gauge, and
it is clear, therefore, that this configuration is the same as the one we studied in section 2,
see Fig.(\ref{zevolv}).
 In particular, one of the parameters of the solution, for example $\om$ can be determined from
the periodicity condition for $z$ which takes the same form as equation (\ref{period}). The velocity  $v$ can then be expressed in terms of $p$ by using the boundary condition (\ref{bc}) for $\p$ and takes the following form
\bea\la{pwscg}
p = \int_{-r}^{r} d\s\, \p' = 2\int_{z_{min}}^{z_{max}}\, {\rm d}z\, {\vp'\ov |z'|}
\,.
\eea
Since ${\vp'\ov |z'|} =p_z$ for $a=0$, see (\ref{pzu}), the change of $\p$ is just equal
to the world-sheet momentum $\pws$ in the $a=0$ uniform gauge. This is what one should expect
because we supplemented the conformal gauge by the condition $t=\tau$.

Finally, the charge $J$ is found by using equation (\ref{Jcg})
 \bea\la{Jcg2}
J = {\sqrt{\lambda}\over \pi}\int_{z_{min}}^{z_{max}}\, {\rm d}z\, {\om(1-z^2)(1-v\vp')\ov |z'|}\,.
\eea
All these integrals can be easily computed in terms of elliptic functions by using formulas from
Appendix B. Computing the integrals, we find that the soliton energy $E-J$, charge $J$ and  momentum $p$ are given by \emph{exactly} the same formulas (\ref{ef}), (\ref{Jf}) and (\ref{relf})
as in the $a=0$ uniform gauge. Therefore, the dispersion relation in the conformal gauge has the same form as the one in the $a=0$ gauge, and
the leading and subleading exponential corrections to $E-J$ are given by
\bea\label{gecg} E-J&=&\frac{\sqrt{\lambda}}{\pi}\sin\frac{p}{2}
\Big[1-\frac{4}{e^2}\sin^2\frac{p}{2}~ e^{-\j}-\\ \nonumber
&&-\frac{4}{e^4}\sin^2\frac{p}{2}~\Big( \j^2(1+\cos
p)+2\j(2+3\cos p) +7+6\cos p
\Big)e^{-2\j}+\cdots \Big]\, , \nonumber \eea
where the effective length which measures the magnitude of the correction is
\bea\la{calJcg}
\j =\frac{2\pi J}{\sqrt\l\sin\frac{p}{2}} \,.
\eea
The ``half-GKP" solution again corresponds to the limit
$v\to 0$ or $p\to \pi$. The finite $J$ correction to the dispersion
formula is of course given by the same equation (\ref{halfGKPen3}).
\begin{figure}[t]
\begin{center}
\includegraphics*[width=.4\textwidth]{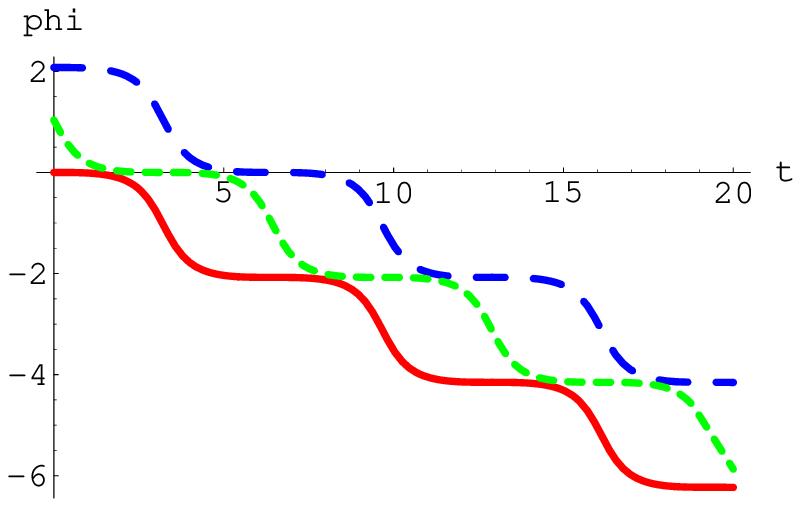}
\includegraphics*[width=.4\textwidth]{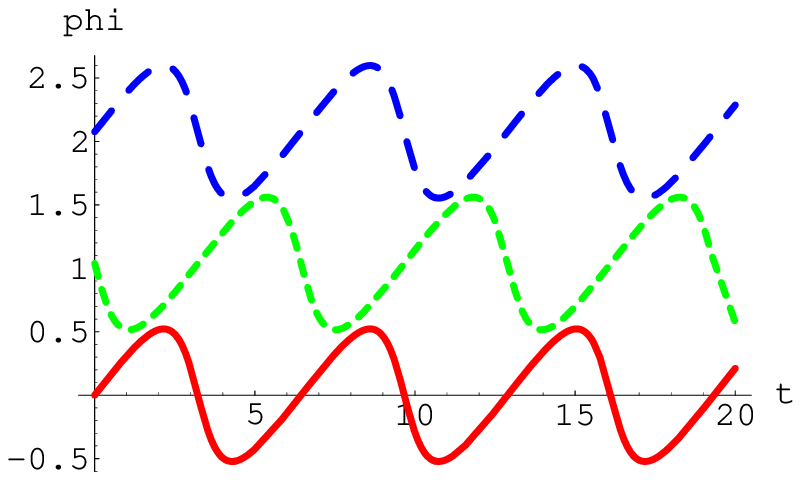}
\end{center}
\caption{\label{eff}Left plot: A time evolution of the end  and middle points of
finite $J$ magnon in $\tilde{\varphi}$ direction. Red and blue lines are string end points, while green line is the middle point of the string. Right plot: Motion of end and middle point  after subtraction of the center of mass motion.
Both plots are made for $\omega = 1.3$ and $v=0.4$
}
\end{figure}

\begin{figure}[t]
\begin{center}
\includegraphics*[width=.6\textwidth]{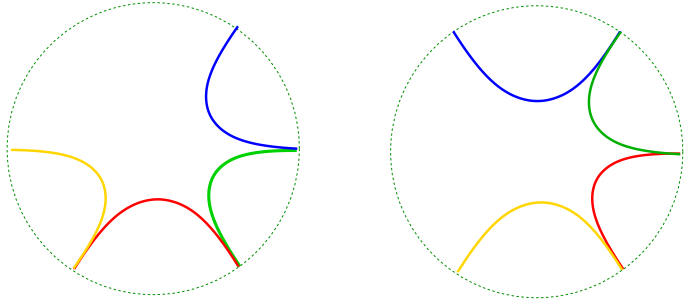}
\end{center}
\caption{\label{star1}Plot of the superposition of $N$ magnons with equal world-sheet momenta. 
The string is nonrigid.
All the individual magnons are hopping in the same direction and with the same velocity. The left picture shows the configuration at $t=0$, and the right one shows the configuration after one hop. }
\end{figure}

\begin{figure}[t]
\begin{center}
\includegraphics*[width=.29\textwidth]{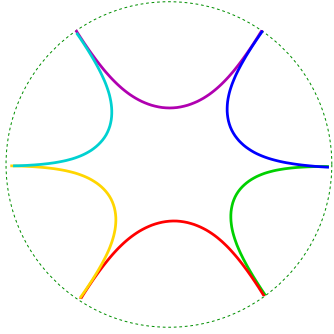}
\end{center}
\caption{\label{star}Plot of the $N$-magnon case: $N \Delta \phi = 2 \pi$. This is
a legitimate closed string configuration. Closed string is
\emph{rigid}! We do not see hopping of the individual magnons any
more because there are no end-points.  }
\end{figure}
\medskip

\begin{figure}[t]
\begin{center}
\psfrag{1/om1}{\smaller\smaller $\sin \theta = \frac{1}{\omega}$}
\psfrag{v/om1}{\smaller\smaller $\sin \theta = v$}
\psfrag{Equator}{\smaller Equator}
\psfrag{Dphi}{$\Delta \varphi$}
\includegraphics*[width=.45\textwidth]{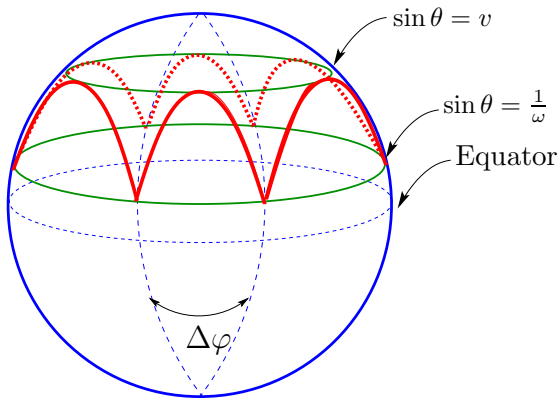}
\end{center}
\caption{\label{Dphi-plot2}Plot of the $N$-magnon configuration: $N \Delta \phi = 2 \pi$. }
\end{figure}

In the conformal gauge case, it is not difficult to write down an
explicit solution of equation (\ref{dth}) by using
Jacobi elliptic functions
\bea
\label{sol2}
z=\frac{\sqrt{1-v^2}}{\om\sqrt\eta}{\rm dn}\Big(\frac{1}{\sqrt\eta}{\s-v\tau\ov\sqrt{1-v^2}}, \eta\Big)\, , \eea
where
\begin{equation}\nonumber
\eta=\frac{1-\om^2v^2}{\om^2(1-v^2)}
\end{equation}
is an elliptic modulus which is determined by the periodicity condition. This formula allows one to
understand easily the target  space-time evolution of the soliton, see Fig.\ref{zprofile} and the simulations  at~\url{http://www.aei.mpg.de/~peekas/magnons/}.

Let us discuss the geometry  of the finite $J$ magnon solution.
The solution in the $z$ direction is clearly periodic,
since it is proportional to Jacobi elliptic function $\rm{dn}$  (\ref{sol2}).
The motion of the end points ($\s= \pm r$) and of the middle point ($\s=0$)
is periodic with the period $T = 2r/v$, and is depicted as a function of time on figure (\ref{ez}).

We see that as $\omega \rightarrow 1$ (corresponding to the limit
of infinite $J$), the period diverges, corresponding to the fact
that it takes infinite amount of target space-time for the soliton
to propagate from one end of the string to another, given the fact
that ``effective'' string length is infinite in this limit.

We also see that both $z$ and $z'$ are periodic functions,
but $z'$ vanishes only at $t = n T/2$. Hence, unlike the infinite $J$ giant
magnon, our string does not satisfy Neumann boundary conditions
for all times.

Motion in the $\vp$ direction is also
non-trival as can be seen by (numerically) integrating equation
(\ref{dphi}). The time evolution of integrated expression for $\varphi$ is shown on
the left plot of figure (\ref{ep}). We see that in addition to the global motion $\omega t$, for finite $J$ configuration, motion in $\varphi$ direction also contributes to the center of mass motion. Subtracting this contribution, we obtain the periodic motion depicted on the right hand side plot of figure (\ref{eff}).

\medskip

Let us also mention that in the case when $p = {2\pi\ov N}$ we can
build a closed string configuration by gluing $N$ finite $J$
solitons, see Fig.(\ref{star1}), (\ref{star}) and (\ref{Dphi-plot2}). The
resulting configuration carries the charge $NJ$ and was studied in
\cite{Ryang}.

%%%%%%%%%%%%%%%%%%%%%%%%%%%%%%%%%%%%%%%%%%%%%%%%%%%%%%%%%%%%%%%%%
\section{Two-spin giant magnon}
%%%%%%%%%%%%%%%%%%%%%%%%%%%%%%%%%%%%%%%%%%%%%%%%%%%%%%%%%%%%%%%%%%

In this section we show that a 2-spin giant magnon configuration
recently discussed in \cite{Dorey} can be easily obtained by ``boosting''\footnote{Here word ``boosting'' is used in the loose sense,  as boost symmetry on a world-sheet is broken  by gauge fixing.}
a giant magnon in an orthogonal direction in the same way as the usual
2-spin folded string solutions were found \cite{FT}. For simplicity we
restrict our consideration to the conformal gauge and infinite $J$
case but similar solutions exist also in a unitary gauge and for
finite $J$. The finite $J$ 2-spin solution in the conformal gauge is
briefly discussed in Appendix D.

The action for strings in ${\mathbb R}\times {\rm S}^3$ is the sum
of the action (\ref{Scg}) for strings in ${\mathbb R}\times {\rm
S}^2$ and a term depending on the angle $\a$ parametrising the
second isometry direction of ${\rm S}^3$: \bea \la{Scg2s} S =
-{\sqrt{\lambda}\over 4\pi}\int_{-r}^{ r}\, {\rm d}\s{\rm d}\tau\,
\left({\pa_\mu z \pa_\mu z\ov 1-z^2} + (1-z^2)\pa_\mu\p \pa_\mu\p
+ z^2 \pa_\mu \a\pa_\mu\a\right)\,, \eea and we also impose the
Virasoro constraints \bea\la{Vcs2} &&{\dz^2+z'^2\ov 1-z^2}
+(1-z^2)\left(
\dph^2 + \p'^2\right) +z^2\left( \da^2 + \alpha'^2\right) = 1\,,\\
&& {\dz z'\ov 1-z^2} +(1-z^2)\dph\p'+ z^2\da\pra= 0\,.  \eea The
two charges $J_1\equiv J$ and $J_2$ corresponding to shifts of
$\p$ and $\a$ are \bea\la{Jcg1} J &=& {\sqrt{\lambda}\over
2\pi}\int_{-r}^{ r}\, {\rm d}\s\, (1-z^2) \dph \,,\\\nonumber J_2
&=& {\sqrt{\lambda}\over 2\pi}\int_{-r}^{ r}\, {\rm d}\s\, z^2
\da\,. \eea In the infinite $J$ case we can look for soliton
solutions of the form \bea\la{ansatzs2} z=z(\s-v\tau)\,,\quad \p
=\vp(\s-v\tau) + \tau\,,\quad \a=\nu \tau - \nu v \s\,.  \eea The
first term $\nu \tau$ in the ansatz for $\a$ describes the motion
along the circle parametrized by $\a$. It appears because we boost
the infinite $J$ giant magnon in the direction parametrized by
$\a$. One can easily check, however, that the equation of motion
for $\a$ forces us to add the second term proportional to $\s$.

Substituting the ansatz (\ref{ansatz}) in the Virasoro constraints (\ref{Vcs2}),  we get
\bea\la{pp2s}
\vp' &=& {v\ov 1-v^2}{z^2\ov 1-z^2}\,,\\\la{zp2s}
z'^2 &=&z^2\,{(1-v^2)(1 -\nu^2(1-v^2)) - (1-\nu^2(1-v^2)^2)z^2 \ov (1-v^2)^2}
\,.
\eea
The solution to equation (\ref{zp2s}) is
\bea\la{sols2}
z={\zeta\ov \cosh(\g(\s-v\tau))}\,,
\eea
where the parameters $\zeta$ and $\g$ are defined as follows
$$
\zeta = \sqrt{{(1-v^2)(1 -\nu^2(1-v^2))\ov 1-\nu^2(1-v^2)^2}}\,,\quad 
\g=\sqrt{{1-\nu^2(1-v^2)\ov 1-v^2}}\,.
$$
Note that the parameters satisfy the identity
$
{\g\sqrt{1-\zeta^2}\ov\zeta} = {1-v^2\ov v^2}\, .
$

The solution (\ref{sols2}) can be easily used to compute $p$,
$J_2$ and $E-J$. We obtain \bea \nonumber p&=& 2\arcsin\zeta\,,\\
\nonumber J_2&=& {\sqrt{\lambda}\over \pi} \nu {\zeta^2\ov
\g}\,,\\ \nonumber E-J&=& {\sqrt{\lambda}\over \pi}{\zeta^2\ov \g
(1-v^2)}\,. \eea Finally, taking into account the identity $ \nu^2
+ {\g^2\ov\zeta^2}={1\ov (1-v^2)^2}\,, $ we obtain the dispersion
relation for the 2-spin magnon \bea E-J = \sqrt{J_2^2 +
{\lambda\over \pi^2}\sin^2\frac{p}{2}}\,. \eea The solution and
the dispersion relations coincide with the ones found in
\cite{Dorey} by using a rather non-trivial relation of the string
sigma model on ${\mathbb R}\times {\rm S}^3$ with the complex
sine-Gordon equation. Our approach
 can be easily  generalised to find a 3-spin giant magnon configuration.

%%%%%%%%%%%%%%%%%%%%%%%%%%%%%%%%%%%%%%%%%%%%%%%%%%%%%%%%%%%%%%%%%%%%%%%
%%%%%%%%%%%%%%%%%%%%%%%%%%%%%%%%%%%%%%%%%%%%%%%%%%%%%%%%%%%%%%%%%%%%%%%
\section*{Acknowledgments }
%%%%%%%%%%%%%%%%%%%%%%%%%%%

We are  grateful to R. Janik, J. Maldacena, K. Peeters, J. Plefka,
M. Staudacher and A. Tseytlin for useful discussions. 
Special thanks to K. Peeters for helping us to make 
the giant worms animation.
This work was supported in part by the grant {\it Superstring Theory}
(MRTN-CT-2004-512194). 
The work of
G.~A. was supported in part by the RFBI grant N05-01-00758, by NWO grant 047017015 
and by the INTAS
contract 03-51-6346.

%%%%%%%%%%%%%%%%%%%%%%%%%%%%%%%%%%%%%%%%%%%%%%%%%%%%%%%%%%%%%%%%%%%%%%%
%%%%%%%%%%%%%%%%%%%%%%%%%%%%%%%%%%%%%%%%%%%%%%%%%%%%%%%%%%%%%%%%%%%%%%%

\appendix

%%%%%%%%%%%%%%%%%%%%%%%%%%%%%%%%%%%%%%%%%%%%%%%%%%
\section{Some explicit formulas}
%%%%%%%%%%%%%%%%%%%%%%%%%%%%%%%%%%%%%%%%%%%%%%%%%%
Here we present explicit expressions for the formulas from section 3
and specify them to the three simplest cases $a=0,1/2,1$.

The density of the gauge-fixed Hamiltonian $\H$ appearing in (\ref{Su1}) as a
function of the coordinate $z$ and the momentum $p_z$ canonically conjugate to $z$ is
\bea
\la{denHu1}
\H &=& -\frac{1-(1-a)z^2}{1-2a-(1-a)^2z^2} \\\nonumber
&~& + \frac{\sqrt{1+\left(1 - z^2\right) \left(1-2a-(1 - a)^2 z^2\right)p_z^2}
\sqrt{1-z^2 + \left(1-2a-(1-a)^2z^2\right) z'^2}}{1-2a-(1-a)^2z^2 }\,,
\eea
The density of the Hamiltonian (\ref{denHu1}) for the three simplest cases:
\bea
\la{denHa0} \nonumber
a=0:\quad \H &=&  -1+\sqrt{\frac{1+z'^2}{1 - z^2}}\sqrt{1+p_z^2 \left(1 - z^2\right)^2} \,,
 \\\la{denHa12} \nonumber
a={1\ov 2}:\quad \H &=& -2 + \frac{4}{z^2} -\frac{1}{z^2}
\sqrt{4(1- z^2)-z^2z'^2}\sqrt{4-p_z^2  z^2\left(1-z^2\right)}\,,
 \\\la{denHa1}\nonumber
a=1:\quad \H &=& 1 -\sqrt{1-z^2 - \left(z'\right)^2}\sqrt{1-\left(1-z^2\right) p_z^2 }\,.
\eea
Solving the equation of motion for $p_z$ following from the action (\ref{Su1}),
we determine the momentum as a function of $\dz$ and $z$
\bea\la{pza}
p_z=\frac{\dz}{\sqrt{\left(1-z^2\right)} \sqrt{\left(1-z^2\right)^2-\left(1-2 a-(1-a)^2 z^2\right)
   \left(\dz^2-\left(1-z^2\right) \left(z'\right)^2\right)}}\,.
\eea
The momentum $p_z$ as a function of $\dz$ and $z$ for the three simplest cases:
\bea\la{pza0}\nonumber
a=0:\quad p_z&=& {\dz \ov (1-z^2)\sqrt{1-z^2-\dz^2 + (1-z^2) z'^2}}  \,,
 \\\la{pza12} \nonumber
a={1\ov 2}:\quad p_z&=&  \frac{2 \dz}{\sqrt{1 - z^2}
\sqrt{4 \left(1 - z^2\right)^2+ z^2\left(\dz^2 - \left(1-z^2\right) z'^2\right) }}\,,
 \\\la{pza1} \nonumber
a=1:\quad p_z&=&  \frac{\dz}{\sqrt{1 - z^2}\sqrt{ \left(1 -
z^2\right)^2 + \dz^2-\left(1 - z^2\right)z'^2 }} \,.
\eea
Substituting the solution (\ref{pza}) into the action (\ref{Su1}),
we obtain the action in the Lagrangian form
\bea\la{Su2}
&&S ={\sqrt{\lambda}\over 2\pi}
\int_{- r}^{ r}\, {\rm d}\s{\rm d}\tau\, \left(
\frac{1 - (1 - a) z^2}{1 - 2 a -(1 - a)^2 z^2} \right.\\\nonumber
&&~~~~~~~~~~~~~~~~\left.
-\frac{\sqrt{\left(1 - z^2\right)^2 - \left( 1- 2 a -(1 - a)^2 z^2\right)
\left(\dz^2 - \left(1 - z^2\right) z'^2\right)}}{\sqrt{1 - z^2}
\left( 1- 2 a -(1 - a)^2 z^2\right)} \
\right)\,.
\eea
The action (\ref{Su1}) in the Lagrangian form for the three simplest cases:
\bea\la{Su2a0}\nonumber
a=0:\quad &S& ={\sqrt{\lambda}\over 2\pi}
\int_{- r}^{ r}\, {\rm d}\s{\rm d}\tau\, \left(
1-{\sqrt{1-z^2-\dz^2 + (1-z^2) z'^2}\ov 1-z^2}
\right)\,, \\\la{Su2a12}\nonumber
a={1\ov 2}:\quad &S& ={\sqrt{\lambda}\over 2\pi}
\int_{- r}^{ r}\, {\rm d}\s{\rm d}\tau\, \left(
2 -\frac{4}{z^2}
+ \frac{2 \sqrt{4 \left(1 - z^2\right)^2+z^2\left(\dz^2 -\left(1 - z^2\right) z'^2\right)}}
{z^2 \sqrt{1 - z^2}}\
\right)\,, \\\la{Su2a1} \nonumber
a=1:\quad &S& ={\sqrt{\lambda}\over 2\pi}
\int_{- r}^{ r}\, {\rm d}\s{\rm d}\tau\, \left(-1 +
\frac{\sqrt{\left(1 - z^2\right)^2  + \dz^2- \left(1 -z^2\right)z'^2}}{\sqrt{1 - z^2}}
\right)\,.
\eea
Substituting the ansatz (\ref{ansatz}) into the action (\ref{Su2}),
we get the following Langrangian of the reduced model
\bea\la{lred}\nonumber
L_{red} = \frac{1 - (1 - a) z^2}{1 - 2 a -(1 - a)^2 z^2}
-\frac{\sqrt{\left(1 - z^2\right)^2 + \left( 1- 2 a -(1 - a)^2 z^2\right)
\left(1-v^2 - z^2\right) z'^2}}{\sqrt{1 - z^2}
\left( 1- 2 a -(1 - a)^2 z^2\right)} \,. \, \,
\eea
The Hamultonian of the reduced one-dimensional model is
\bea\nonumber
&&H_{red} = \pi_z z' - L_{red}= -\frac{1 - (1 - a) z^2}{ 1- 2 a -(1 - a)^2 z^2} \\\nonumber
&&~~+
\frac{\left(1 - z^2\right)^{3/2}}{\left( 1- 2 a -(1 - a)^2 z^2\right)
\sqrt{\left(1 - z^2\right)^2 + \left( 1- 2 a -(1 - a)^2 z^2\right)
\left(1-v^2 - z^2\right) z'^2}
}\,.
\eea
$z'^2$ as a function of $z$ for the three simplest cases:
\bea\la{zpa0}\nonumber
a=0:\quad &z'^2& = \frac{1-\om^2 v^2}{1-v^2 - z^2} -1\,,
\\\la{zpa12} \nonumber
a={1\ov 2}:\quad &z'^2& =  \left({2 \left(1 - z^2\right)\ov 1+\om -\om z^2}\right)^2{1-\om^2 + \om ^2z^2 \ov 1-v^2 - z^2}\,,
\\\la{zpa1} \nonumber
a=1:\quad &z'^2& = \left(1 - z^2\right)^2{1-\om^2 + \om ^2z^2 \ov 1-v^2 - z^2}
\,.
\eea
$\H/|z'|$ as a function of $z$ for the three simplest cases:
\bea\la{ha0}\nonumber
a=0:\quad {\H\ov |z'|}& =& \frac{\om z^2 -(\om-1)(1+\om v^2)}{\sqrt{1-v^2 - z^2}\sqrt{1-\om^2 + \om ^2z^2 }} \,,
\\\la{ha12} \nonumber
a={1\ov 2}:\quad {\H\ov |z'|}& =&
\frac{-\om(\om + 1) z^4 + \left(2\om^2+\om-1+\om(\om-1) v^2\right) z^2 - (\om^2-1)(1+v^2) }
{(\om +1)(1-z^2)\sqrt{1-v^2 - z^2}\sqrt{1-\om^2 + \om ^2z^2 }}\,,
\\\la{ha1} \nonumber
a=1:\quad {\H\ov |z'|}& = &\frac{-\om^2 z^4 + \om\left(2\om-1\right) z^2 - (\om-1)(\om+v^2) }
{\om (1-z^2)\sqrt{1-v^2 - z^2}\sqrt{1-\om^2 + \om ^2z^2}}
\,.
\eea

%%%%%%%%%%%%%%%%%%%%%%%%%%%%%%%%%%%%%%%%%%%%%%%%%%%%%%%%%%%%%%
\section{Finite $J$ corrections to the dispersion relation}
%%%%%%%%%%%%%%%%%%%%%%%%%%%%%%%%%%%%%%%%%%%%%%%%%%%%%%%%%%%%%%%

Here we will outline the derivation of the leading finite $J$
correction to the dispersion formula in the uniform gauge.

To find explicit expressions for the energy,
charge and world-sheet momentum we need to use the following formulas
\bea \la{integrals}
I_1=\int_{z_{min}}^{z_{max}}\, {\rm d}z \frac{1}{\sqrt{z^2 - z_{min}^2}\sqrt{z_{max}^2-z^2}}
=\frac{1}{z_{max}}\K(\eta)\,,~~~~~~~~~~~~~~~~~~~~~~~~~~~~~~~~~~~~~~~~&&\\\nonumber
I_2=\int_{z_{min}}^{z_{max}}\, {\rm d}z \frac{z^2}{\sqrt{z^2 - z_{min}^2}\sqrt{z_{max}^2-z^2}}
=z_{max}\E(\eta)\,,~~~~~~~~~~~~~~~~~~~~~~~~~~~~~~~~~~~~~~~~~&&\\
\nonumber I_3=\int_{z_{min}}^{z_{max}} {\rm d}z
\frac{1}{(1-z^2)\sqrt{z^2 -
z_{min}^2}\sqrt{z_{max}^2-z^2}}=\frac{1}{z_{max}(1-z_{max}^2)}\Pi\Big(\frac{z_{max}^2-z_{min}^2}
{z_{max}^2-1},\eta\Big)&&
\eea 
Here $\eta$ is the elliptic modulus defined as
$$
\eta=1-\frac{z_{min}^2}{z_{max}^2}\, .
$$

\vskip 0.5cm

Computing the half-period $r$ of the solution by using the formulas (\ref{integrals}), we obtain
\bea\la{rf}
r=(1-a)\sqrt{1-v^2}\left(\K(\eta)-\E(\eta)\right)
+{a\ov \om\sqrt{1-v^2}}\left(\K(\eta)
-\Pi\Big(\frac{v^2-1}{v^2}\eta,\eta\Big)\right) \,, \,
\eea
where the elliptic modulus is related to the parameters of the solution as follows
\bea\la{mo}
\eta=1-{z_{min}^2\ov z_{max}^2}=\frac{1-\om^2v^2}{(1-v^2)\om^2}\, .
\eea
One can easily see from this expression that as $\om\to 1$, the modulus approaches 1,
and the period goes to infinity in accord with the discussion in section 3.

For the energy of the soliton (\ref{energya}) we find the following result
\bea\la{ef}
E-J&=&\frac{\sqrt{\lambda}}{\pi}\Big[ \sqrt{1-v^2} \E(\eta) \\
\nonumber &&~~~~~~-\left(\om-1\right)\frac{\Big(1+\om v^2 +a(\om-1-\om v^2)\Big)
\K(\eta)
+a\Pi\Big(\frac{v^2-1}{v^2}\eta,\eta\Big)}{\om(1-a+a\om)\sqrt{1-v^2}}\Big]\,.
  \eea
As $\om\to 1$, $\eta\to 1$, the second $a$-dependent line drops out and
one is left with the first term which gives the dispersion relation (\ref{HMdis}) of
the giant magnon.

To analyze the $a$-dependence of the finite $J$ dispersion relation, it is necessary to express
the soliton energy as a function of $\pws$, $J$ and $a$. It can be done by means of the formula
$$
J = P_+ - a(E-J)\,,
$$
which allows us to find $J$ as a function of the parameters of the solution
\bea\la{Jf}
J=\frac{\sqrt{\lambda}}{\pi}\Big[\sqrt{1-v^2}\left(\K(\eta)-\E(\eta)\right)
+\frac{a\om^2  v^2
\K(\eta) - a \Pi\Big(\frac{v^2-1}{v^2}\eta,\eta\Big)}{\om(1-a+a\om)\sqrt{1-v^2}}
\Big]\,.
\eea
Finally, for the world-sheet momentum $\pws$ the following expression is found
\bea\la{pf}
\pws=-\frac{2\om v}{(1-a+a\om)\sqrt{1-v^2}}\K(\eta)+\frac{2
 \Pi\Big(\frac{v^2-1}{v^2}\eta,\eta\Big)}{\om(1-a+a\om)\sqrt{1-v^2}}\,
. \eea
Comparing equation.(\ref{Jf}) and (\ref{pf}), we see that the following simple relation holds
\bea\la{relf}
J + \frac{\sqrt{\lambda}}{\pi} a v {\pws\ov 2} =
\frac{\sqrt{\lambda}}{\pi}\sqrt{1-v^2}\left(\K(\eta)-\E(\eta)\right)\,
.
\eea
This relation can be used to express the modulus $\eta$ in terms of $J$, $\pws$, $v$ and $a$. Then,
the velocity $v$ can be found as a function of $J$, $\pws$ and $a$ from equation (\ref{pf}).
This gives the soliton energy (\ref{ef}) as a function of $J$, $\pws$ and $a$, that is the finite
$J$ dispersion relation. It is obvious that there is no simple analytic expression for the dispersion
relation. It is possible, however, to analyze it for large values of the charge $J$.

To this end, expressing from equation (\ref{pf}) elliptic
integral of the the third kind via the momentum $\pws$ and
substituting in into expression (3.18), we obtain the following
formula
\bea\label{energie} && E-J=\frac{\sqrt{\lambda}}{2\pi}\Big[
av\pws\Big(1-\frac{1}{\sqrt{\eta+v^2(1-\eta)}}\Big)\\ \nonumber
&&~~~~~~~~~~~~~~~~~~~~~~~~~+2\sqrt{1-v^2}
\Big(\E(\eta)-\Big(1-\frac{\eta}{\sqrt{\eta+v^2(1-\eta)}}\Big)\K(\eta)\Big)
\Big]\, .
 \eea

\vskip 0.5cm

To consider the asymptotic expansion $J\to \infty$ it is
convenient to introduce the variable $\eps=1-\eta$. Obviously,
$\eps\to 0$ as $\eta\to 1$. We also express the variable $\om$ via
$\eps$ and $v$:
$$
\om=\frac{1}{\sqrt{1-(1-v^2)\eps}}\,.
$$
Furthermore, it is convenient to transform the elliptic integral
of the third kind with modulus $\eta$ to the integral with the
complementary modulus $\eps$. The relevant transformation formula
is
\bea \nonumber
\Pi\Big(\frac{v^2-1}{v^2}\eta,\eta\Big)
&=&\frac{1}{(1-(1-v^2)\eps)\K(\eps)}\Big[ \frac{\pi
v}{2}\sqrt{1-v^2}\sqrt{1-(1-v^2)\eps}~{\rm
F}\Big(\arcsin\sqrt{1-v^2},\eps\Big)\\
\nonumber
&+&\K(1-\eps)\Big((1-(1-v^2)\eps)\K(\eps)-(1-\eps)(1-v^2)\Pi\Big(\frac{\eps
v^2 }{1-(1-v^2)\eps},\eps\Big)\Big) \Big]\, ,
\eea
where
${\rm F}(\varphi,\eps)$ is the standard incomplete elliptic integral of
the first kind. With this formula at hand the expression for the
world-sheet momentum can be cast into the form
\bea\nonumber
\pws&=&\frac{1+\sqrt{1-(1-v^2)\eps}}{v(1-(1-a)(1-v^2)\eps+\sqrt{1-(1-v^2)\eps})\K(\eps)}\times
\\\nonumber
&~&\Big[\pi v \sqrt{1-(1-v^2)\eps} ~{\rm
F}\Big(\arcsin\sqrt{1-v^2},\eps\Big)\\
\label{mom}
&~&~~~~~~~~+2\sqrt{1-v^2}(1-\eps)\K(1-\eps)\Big(\K(\eps)-\Pi\Big(\frac{\eps
v^2 }{1-(1-v^2)\eps},\eps\Big)\Big)\Big]\, .
 \eea
To develop an asymptotic expansion $\eps\to 0$ one can use the
following formula
\bea
\Pi\Big(\frac{\eps v^2
}{1-(1-v^2)\eps},\eps\Big)=\frac{\pi}{2}&+ &
\frac{\pi}{8}(1+2v^2)\eps
 + \frac{\pi}{128}(9+44v^2-8v^4)\eps^2 \nonumber \\\nonumber
&+&\frac{\pi}{512}(25+206 v^2-72 v^4+16 v^6)\eps^3+\cdots \, ,
\eea
where we have assumed for the moment that $v$ is kept constant
($\eps$-independent).

We treat equation (\ref{mom}) as an equation for $v$ regarded as the
function of $\pws$ and the modulus $\eps$. One can see that this
equation can be solved by assuming the following expansion for
$v$:
$$
v=\cos\frac{\pws}{2}+\sum_{k=1}^{\infty}\sum_{m=0}^k
c_{km}(\pws)(\log\eps)^m\eps^k \, .
$$
For the leading coefficients we find from
equation (\ref{pf})\bea\nonumber
c_{10}&=&-\frac{1}{4}\sin^2\frac{\pws}{2}\Big(\cos\frac{\pws}{2}(1+\log
16 )+a\pws\sin\frac{\pws}{2}\Big)\, ,\\\nonumber
c_{11}&=&\frac{1}{4}\sin^2\frac{\pws}{2}\cos\frac{\pws}{2}\, .
 \eea
Further computation of the subleading coefficients give
\bea\nonumber c_{20}&=&-\frac{1}{2^6}\sin^2\frac{\pws}{2}
\Big[\cos\frac{\pws}{2}\Big( 1+5a^2\pws^2+40\log 2 +16\log^2 2 \\
\nonumber
&&~~~~~~~~~~~~~~~~~~~~+\cos \pws(-5a^2\pws^2+2+16\log 2+80\log^2 2) \Big)+\\
 &&~~~~~~~~~~~~~ - 4a\pws\sin\frac{\pws}{2}\Big(3+\cos \pws+10\cos
\pws\log 2+6\log 2 \Big)\Big]\, , \nonumber \\
%%%%%%%%%%%%%%%%%%%%%%%%%%%%%%%%%%%%%%%%%%%%%%%%%%%%%%%%%%%%%%%%%
\nonumber
 c_{21}&=&\frac{1}{2^7} \Big[ \cos\frac{\pws}{2}(5+4\log
2)+\cos\frac{3\pws}{2}(-4+6\log 2)-\cos\frac{5\pws}{2}(1+10\log
2)+\\\nonumber
&&~~~~~~~~~~~~~~~~~~+4a\pws(3+5\cos \pws)\sin^3\frac{\pws}{2}\Big] \, ,\\
\nonumber
c_{22}&=&-\frac{1}{2^7}\sin^2\frac{\pws}{2}\Big[7\cos\frac{\pws}{2}+5\cos\frac{3\pws}{2}\Big]\,
.
 \eea
Now substituting  the expansion for $v$ into equation (\ref{relf}) we
can determine the dependence of the modulus $\eps$ on $\pws$ and
$\cj=\frac{2\pi J}{\sqrt{\lambda}}$. We find that the modulus is
expandable into series
$$
\eps=\frac{16}{e^2}e^{-\j}\sum_{m=0}^{\infty} a_m(\pws)e^{-m\j} \,,
$$
where $$\j =\frac{2\pi J}{\sqrt\l\sin\frac{\pws}{2}} +a\pws\cot\frac{\pws}{2}\,.$$
For the first two leading terms we find \bea\nonumber
a_0(\pws,\cj)&=&1 \\
a_1(\pws,\cj)&=&\frac{2}{e^2\sin^2\frac{\pws}{2}}\Big[
2(-1+a^2\pws^2+\cos \pws+ a\pws\sin \pws
)\nonumber \\\nonumber
 &+&{\cj\ov 2}(8a\pws\cos\frac{\pws}{2}-\sin\frac{\pws}{2}
+3\sin\frac{3\pws}{2})
+\cj^2(1+\cos \pws) \Big] \, \, \, \,\, .  \eea

\medskip

\noindent Finally we find the leading and the subleading
corrections to $E-J$:
\bea \label{ge}
E-J&=&\frac{\sqrt{\lambda}}{\pi}\sin\frac{\pws}{2}
\Big[1-\frac{4}{e^2}\sin^2\frac{\pws}{2}~
e^{-\j}-\\ \nonumber
&-&\frac{2}{e^4}\Big((4+4a^2\pws^2-\cos
\pws-3\cos 2\pws +10 a\pws \sin \pws)  \\
&+&
2\cj (4a\pws\cos\frac{\pws}{2}+\sin\frac{\pws}{2}+3\sin\frac{3\pws}{2})+2\cj^2(1+\cos
\pws )\Big)e^{-2\j}+\cdots \Big]\, . \nonumber
\eea
Expressing $\cj$ in terms of $\j$, we obtain equation (\ref{ge3}).

\bigskip

It is of interest to consider the case $v\to 0$ which corresponds
to the finite-$J$ generalization of the ``half-GKP" solution. From
equation.(\ref{relf}) and (\ref{energie}) one can recognize that in
this limit the explicit dependence on $a$ in the expressions for
$E-J$ and $J$ disappears.

Let us now analyze equation (\ref{pf}) in the limit $v\to 0$. Using the
formula
$$
\lim_{v\to
0}\frac{\Pi\Big(\frac{v^2-1}{v^2}\eta,\eta\Big)}{v}=\frac{\pi}{2\sqrt{\eta}}
$$
we find that in this limit
\bea
\pws\to \pi-\frac{a\pi\eps}{1+\sqrt{1-\eps}-(1-a)\eps}\, .
\label{pja}
\eea
Thus, $\pws\to \pi$ for the case $a=0$ only. On the
other hand, for $v=0$ the modulus $\eps$ is a function of $\cj$
only. From expression for $J$ it is easy to find that
$$
\eps=\frac{16}{e^2}e^{-\cj}\Big(1-\frac{4(2+\cj)}{e^2}e^{-\cj}+
\frac{60+4\cj(17+6\cj)}{e^4}e^{-2\cj}+\cdots\Big)\,
.
$$
Substituting this expansion into equation (\ref{energie}) we find
\bea
\label{halfGKPen}
E-J=\frac{\sqrt{\lambda}}{\pi}\Big[1-\frac{4}{e^2}
e^{-\cj}-\frac{4}{e^4}(1-2\cj) e^{-2\cj}+\cdots \Big]\, .
\eea
This
formula gives the first two finite-$J$ corrections to the
``half-GKP" solution. It can be also derived from our general
expansion (\ref{ge}) provided we use the following expansion for
$\pws$:
\bea\nonumber
\pws=\pi-\frac{8\pi a}{e^2}e^{-\cj}
+\frac{32\pi a(-1+2a+\cj)}{e^4}e^{-2\cj}+\cdots\, ,
\label{expansionp}
\eea
which
is a consequence of equation (\ref{pja}). Note that in spite of the
presence of the parameter $a$ in the formula (\ref{ge}) the
expression (\ref{halfGKPen}) is $a$-independent as it should be;
the explicit $a$-dependence is cancelled out upon usage of
equation (\ref{expansionp}).

%%%%%%%%%%%%%%%%%%%%%%%%%%%%%%%%%%%%%%%%%%%%%%%%%%%%%%%%%%%%%%%%%
\section{2-spin giant magnon at finite $J$}
%%%%%%%%%%%%%%%%%%%%%%%%%%%%%%%%%%%%%%%%%%%%%%%%%%%%%%%%%%%%%%%%%%
In this appendix we discuss briefly a finite $J$ 2-spin giant magnon configuration in the conformal gauge.

At finite $J$ we have to impose the periodicity condition for $z$ and $\a$. The periodicity condition
for $\a$ forces us to modify the ansatz (\ref{ansatzs2}) as follows
\bea\la{ansatzd}\nonumber
z=z(\s-v\tau)\,,\quad \p =\vp(\s-v\tau) + \om \tau\,,\quad \a=\nu \tau - \nu v \s +\ta(\s-v\tau)\,.
\eea
The dependence of $\a$ on $\tau$ and $\s$ follows from its equation of motion which takes the following form with the ansatz
\bea\nonumber
\pa_\mu (z^2 \pa_\mu\a)= (v^2-1)\left( z^2 \pta\right)'=0\,.
\eea
Thus the general solution is of the form
\bea\nonumber
 \pta = {C_\a\ov z^2}\ \Rightarrow\  \ta(\s) = C_\a\int_0^\s\, {ds\ov z^2(s)}\,.
\eea
The constant $C_\a$ should be found from the periodicity condition for the angle $\a$
\bea\nonumber
\a(L/2)-\a(-L/2) = 2\pi n = -\nu v L + C_\a\int_{-L/2}^{L/2}\, {ds\ov z^2(s)}\,,
\eea
where $L=2\pi E/\sqrt\l$ and $n$ is an integer which shows the number of times the string winds around the circle parametrized by $\a$..

Substituting the ansatz (\ref{ansatz}) in Virasoro constraints (\ref{Vcs2}), we get
\bea\nonumber
\vp' &=& -{\om v\ov 1-v^2} +    {v -C_\a \nu (1-v^2)^2\ov \om (1-v^2)}   {1\ov 1-z^2}\,\\
z'^2 &=& -{C_\a\ov z^2} +
{(1-\om^2)(\om^2 -v^2) + 2 C_\a \nu v (1-v^2)^2 +C_\a^2(1-v^2)^2(\om^2-\nu^2(1-v^2)^2)
\ov \om^2(1-v^2)^2}\nonumber \\\nonumber
&-&{1-2\om^2+v^2 +\nu^2(1-v^2)^2\ov (1-v^2)^2}z^2 - {\om^2-\nu^2(1-v^2)^2\ov (1-v^2)^2}z^4
\,.
\eea
It seems useful to make an additional change of variables
\bea\nonumber
z = \sqrt{w}\,.
\eea
Then we get
\bea\nonumber
{w'^2\ov 4}  = -C_\a +
{(1-\om^2)(\om^2 -v^2) + 2 C_\a \nu v (1-v^2)^2 +C_\a^2(1-v^2)^2(\om^2-\nu^2(1-v^2)^2)
\ov \om^2(1-v^2)^2}\,w &&\nonumber \\\nonumber
-{1-2\om^2+v^2 +\nu^2(1-v^2)^2\ov (1-v^2)^2}\,w^2 - {\om^2-\nu^2(1-v^2)^2\ov (1-v^2)^2}\,w^3
\,.~~~~~~~~~~~~~~~~~~~~~~~~~~~~~&&
\eea
Even though the equation can be integrated in terms of elliptic functions, the resulting equations
are quite complicated, and we postpone their analysis for elsewhere.

%%%%%%%%%%%%%%%%%%%%%%%%%%%%%%%%%%%%%%%%%%%%%%%%%

\end{document}